\documentclass[namedreferences]{solarphysics}
\usepackage[optionalrh]{spr-sola-addons} 
\usepackage{graphicx}        
\usepackage{color}           
\usepackage{url}             
\usepackage{hyperref}

\begin{document}

\begin{article}

\begin{opening}

\title{The {\it Interface Region Imaging Spectrograph} (IRIS)}

\author{B.~\surname{De Pontieu}$^{1,7}$\sep
         A.M.~\surname{Title}$^{1}$\sep
         J.R.~\surname{Lemen}$^{1}$\sep
         G.D.~\surname{Kushner}$^{1}$\sep
         D.J.~\surname{Akin}$^{1}$\sep
         B.~\surname{Allard}$^{1}$\sep
         T.~\surname{Berger}$^{1,12}$\sep
         P.~\surname{Boerner}$^{1}$\sep
        M.~\surname{Cheung}$^{1}$\sep
        C.~\surname{Chou}$^{1}$\sep
        J.F.~\surname{Drake}$^{1}$\sep
        D.W.~\surname{Duncan}$^{1}$\sep
        S.~\surname{Freeland}$^{1}$\sep
        G.F.~\surname{Heyman}$^{1}$\sep
         C.~\surname{Hoffman}$^{1}$\sep
        N.E.~\surname{Hurlburt}$^{1}$\sep
       R.W.~\surname{Lindgren}$^{1}$\sep
         D.~\surname{Mathur}$^{1}$\sep
       R.~\surname{Rehse}$^{1}$\sep
         D.~\surname{Sabolish}$^{1}$\sep
         R.~\surname{Seguin}$^{1}$\sep
        C.J.~\surname{Schrijver}$^{1}$\sep
         T.D.~\surname{Tarbell}$^{1}$\sep
         J.-P.~\surname{W\"ulser}$^{1}$\sep
         C.J.~\surname{Wolfson}$^{1}$\sep
         C.~\surname{Yanari}$^{1}$\sep
         J.~\surname{Mudge}$^{2}$\sep
         N.~\surname{Nguyen-Phuc}$^{2}$\sep
         R.~\surname{Timmons}$^{2}$\sep
         R.~\surname{van~Bezooijen}$^{2}$\sep
         I.~\surname{Weingrod}$^{2}$\sep
         R.~\surname{Brookner}$^{3}$\sep
         G.~\surname{Butcher}$^{3}$\sep
        B.~\surname{Dougherty}$^{3}$\sep
         J.~\surname{Eder}$^{3}$\sep
         V.~\surname{Knagenhjelm}$^{3}$\sep
         S.~\surname{Larsen}$^{3}$\sep
         D.~\surname{Mansir}$^{3}$\sep
         L.~\surname{Phan}$^{3}$\sep
         P.~\surname{Boyle}$^{3}$\sep
        P.N.~\surname{Cheimets}$^{4}$\sep
         E.E.~\surname{DeLuca}$^{4}$\sep
         L.~\surname{Golub}$^{4}$\sep
        R.~\surname{Gates}$^{4}$\sep
         E.~\surname{Hertz}$^{4}$\sep
         S.~\surname{McKillop}$^{4}$\sep
         S.~\surname{Park}$^{4}$\sep
         T.~\surname{Perry}$^{4}$\sep
        W.A.~\surname{Podgorski}$^{4}$\sep
         K.~\surname{Reeves}$^{4}$\sep
         S.~\surname{Saar}$^{4}$\sep
         P.~\surname{Testa}$^{4}$\sep
         H.~\surname{Tian}$^{4}$\sep
         M.~\surname{Weber}$^{4}$\sep
         C.~\surname{Dunn}$^{5}$\sep
         S.~\surname{Eccles}$^{5}$\sep
         S.A.~\surname{Jaeggli}$^{5}$\sep
         C.C.~\surname{Kankelborg}$^{5}$\sep
         K.~\surname{Mashburn}$^{5}$\sep
         N.~\surname{Pust}$^{5}$\sep
         L.~\surname{Springer}$^{5}$\sep
         R.~\surname{Carvalho}$^{6}$\sep
         L.~\surname{Kleint}$^{6,10}$\sep
         J.~\surname{Marmie}$^{6}$\sep
         E.~\surname{Mazmanian}$^{6}$\sep
         T.M.D.~\surname{Pereira}$^{6,7}$\sep
         S.~\surname{Sawyer}$^{6}$\sep
         J.~\surname{Strong}$^{6}$\sep
         S.P.~\surname{Worden}$^{6}$\sep
         M.~\surname{Carlsson}$^{7}$\sep
         V.H.~\surname{Hansteen}$^{7}$\sep
         J.~\surname{Leenaarts}$^{7}$\sep
         M.~\surname{Wiesmann}$^{7}$\sep
         J.~\surname{Aloise}$^{8}$\sep
         K.-C.~\surname{Chu}$^{8}$\sep
         R.I.~\surname{Bush}$^{8}$\sep
         P.H.~\surname{Scherrer}$^{8}$\sep
%
         P.~\surname{Brekke}$^{9}$\sep
         J.~\surname{Martinez-Sykora}$^{10,1}$\sep
         B.W.~\surname{Lites}$^{11}$\sep
         S.W.~\surname{McIntosh}$^{11}$\sep
        H.~\surname{Uitenbroek}$^{12}$\sep
        T.J.~\surname{Okamoto}$^{13}$\sep
        M.A.~\surname{Gummin}$^{14}$\sep
         G.~\surname{Auker}$^{15}$\sep
         P.~\surname{Jerram}$^{15}$\sep
         P.~\surname{Pool}$^{15}$\sep
         N.~\surname{Waltham}$^{16}$\sep
      }
\runningauthor{B. De Pontieu {\it et al.}}
\runningtitle{{\it Interface Region Imaging Spectrograph}}

   \institute{$^{1}$ Lockheed Martin Solar \& Astrophysics Laboratory,
     Lockheed Martin Advanced Technology Center, Org. A021S,
     Bldg. 252, 3251 Hanover St., Palo Alto, CA 94304, USA 
                     www: \url{http://iris.lmsal.com/} email: \url{bdp@lmsal.com}\\ 
              $^{2}$ Lockheed Martin Advanced Technology Center, Palo
              Alto, 3251 Hanover St., Palo Alto, CA 94304, USA \\
              $^{3}$ Lockheed Martin, 1111 Lockheed Martin Way, Sunnyvale,
              CA 94089, USA
\\
              $^{4}$ Harvard-Smithsonian Astrophysical Observatory, 60 Garden
              Street, Cambridge, MA 02138, USA
                      \\
              $^{5}$ Department of Physics, Montana State University,
              Bozeman, P.O. Box 173840, Bozeman, MT 59717, USA
                     \\
              $^{6}$ NASA Ames Research Center, Moffet Field, CA
              94305, USA
                     \\
              $^{7}$ Institute of Theoretical Astrophysics, University
              of Oslo, P.O. Box 1029 Blindern, Oslo, Norway
                     \\
              $^{8}$ W.W. Hansen Experimental Physics Laboratory,
              Center for Space Science and Astrophysics, Stanford
              University, CA 94305, USA
                     \\
              $^{9}$ Norwegian Space Centre, P.O. Box 113 Skøyen,
              N-0212 Oslo, Norway
                  \\
              $^{10}$ Bay Area Environmental Research Institute,
              596 1st St West, Sonoma, CA 95476, USA
                   \\
             $^{11}$ National Solar Observatory, Sacramento Peak,
              P.O. Box 62 Sunspot, NM 88349-0062, USA
                   \\
             $^{12}$ High Altitude Observatory, National Center for
             Atmospheric Research, P.O. Box 3000, Boulder, CO 80307, USA
                   \\
             $^{13}$ ISAS/JAXA, Sagamihara, Kanagawa 252-5210, Japan
                   \\
           $^{14}$ Alias Aerospace, Inc., 1731 Saint Andrews Court,
              St. Helena, CA 94584, USA
                     \\
              $^{15}$ e2v technologies, 106 Waterhouse Lane,
              Chelmsford, Essex CM1 2QU, UK
                    \\
              $^{16}$ Rutherford Appleton Laboratory, Harwell Business
              Innovation Campus, Didcot, Oxon, OX11 0QX, UK
                    \\
     }

\begin{abstract}
The Interface Region Imaging Spectrograph (IRIS) small explorer
spacecraft provides simultaneous
spectra and images of the photosphere, chromosphere, transition region,
and corona with 0.33\,--\,0.4 arcsec spatial resolution, 2 s temporal
resolution and 1 km~s$^{-1}$ velocity resolution over a field-of-view of up
to 175~arcsec $\times$ 175~arcsec. IRIS was launched into
a Sun-synchronous orbit on 27 June 2013 using a Pegasus-XL rocket and consists of a 19-cm
UV telescope that feeds a slit-based dual-bandpass imaging
spectrograph. IRIS obtains spectra in passbands from 1332-1358\,\AA, 1389-1407\,\AA\ and 2783-2834\,\AA\ including
bright spectral lines formed in the chromosphere (Mg {\sc ii} h 2803\,\AA\ and
Mg {\sc ii} k 2796\,\AA) and transition region (C {\sc ii} 1334/1335\,\AA\ and Si {\sc iv}
1394/1403\,\AA). Slit-jaw images in four different passbands (C {\sc ii}
1330, Si {\sc iv} 1400, Mg {\sc ii} k 2796 and Mg {\sc ii} wing 2830\,\AA) can be taken
simultaneously with spectral rasters that sample regions up to
130 arcsec $\times$ 175 arcsec at a variety of spatial samplings (from
0.33 arcsec and up). IRIS is sensitive to
emission from plasma at temperatures between 5000 K and 10 MK and will advance our understanding of the flow of mass and
energy through an {\sl interface region,} formed by the chromosphere and
transition region, between the photosphere and corona. This highly structured and
dynamic region not only acts as the conduit of all mass and energy
feeding into the corona and solar wind, it also requires an order of
magnitude more energy to heat than the corona and solar wind combined. The IRIS investigation includes a strong
numerical modeling component based on advanced radiative-MHD codes to
facilitate interpretation of observations of this complex
region.
Approximately eight Gbytes of data (after compression) are acquired by IRIS each day and made available
for unrestricted use within a few days of the observation.
\end{abstract}
\keywords{Heating, Chromospheric; Heating, Coronal; Chromosphere, Models; Chromosphere,
  Active; Corona, Active; Magnetic Fields, Chromosphere;
  Instrumentation and Data Management; Spectrum, Ultraviolet}

\end{opening}

\section{Introduction}
     \label{S-Introduction} 

The chromosphere and transition region (TR) form a complex {\sl interface
region} between the solar surface and corona. Almost all of the
mechanical energy that drives solar activity and solar atmospheric
heating is converted into heat and radiation within this interface
region, with only a small amount leaking through to power coronal
heating and drive the solar wind. Understanding the chromosphere and TR 
is a foundational necessity for explaining the corona and
heliosphere. They require a heating rate that is between one and two
orders of magnitude larger than that of the corona. It is also here that we can find information on processes as
diverse as the role of field-line braiding, interaction of the
active-region and network field with the small-scale magnetic carpet,
the interaction of emerging flux with the existing fields, wave
propagation and mode conversion, mass supply to the corona and solar
wind, and signatures of coronal processes through thermal conduction
or energetic particles. 

\begin{figure}    
  \centerline{\includegraphics[width=0.5\textwidth,clip=]{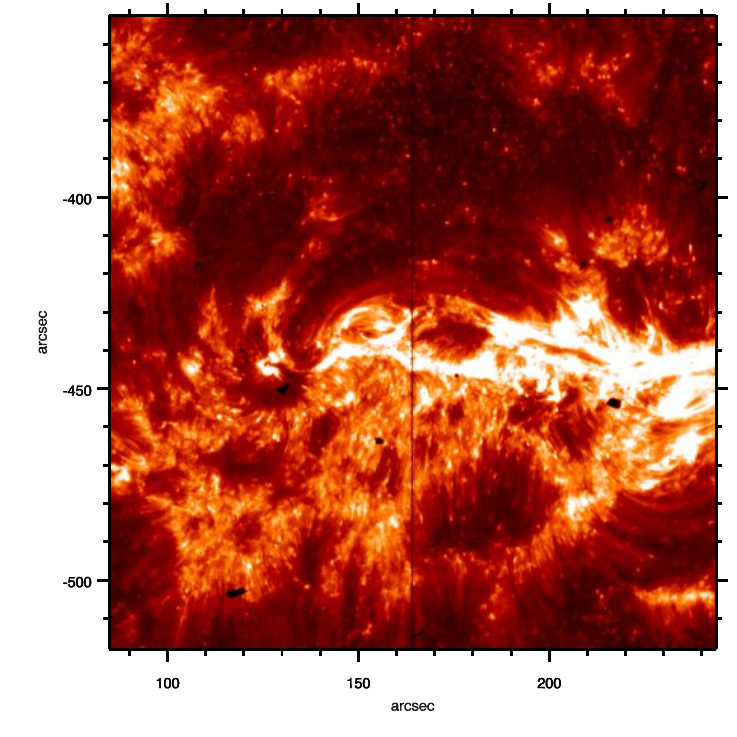}\includegraphics[width=0.5\textwidth,clip=]{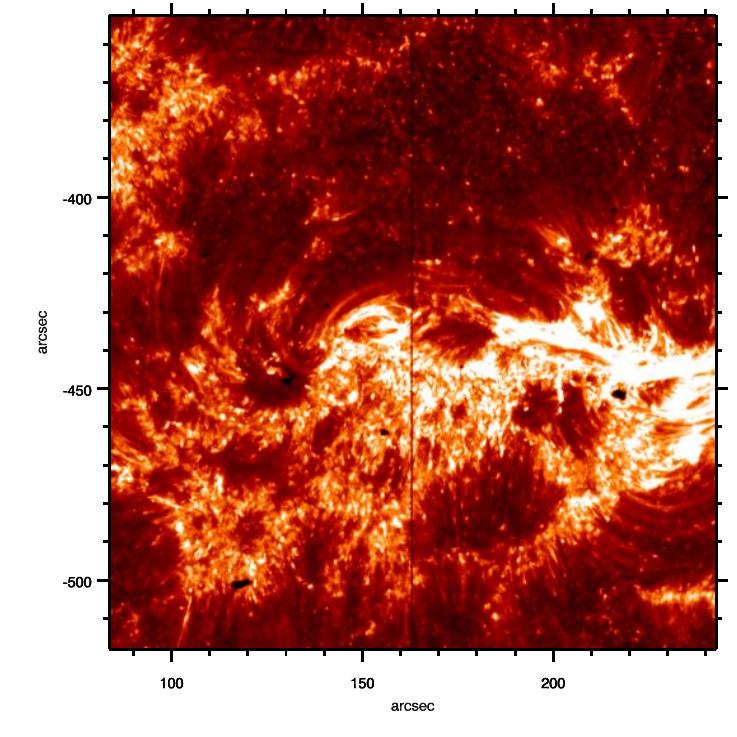}}
              \caption{IRIS slit-jaw images (SJI) 1330\,\AA\ and 1400\,\AA\ of NOAA
                AR 11817 taken on 14 Aug 2013 at 1850 UT. These images are sensitive to plasma of
              10\,000-30\,000 K (1330\,\AA) and 65\,000 K (1400\,\AA) and
              show the upper chromosphere and low transition
              region. Both images also contain contributions from
              continuum that is formed in the low chromosphere. The
              dark vertical line in the middle of the images is the
              location of the slit. Corresponding spectra are shown in
            Figures~\ref{fig1}, \ref{fig2}, \ref{fig3}.}
   \label{fig4}
   \end{figure}

\begin{figure}    
  \centerline{\includegraphics[width=0.5\textwidth,clip=]{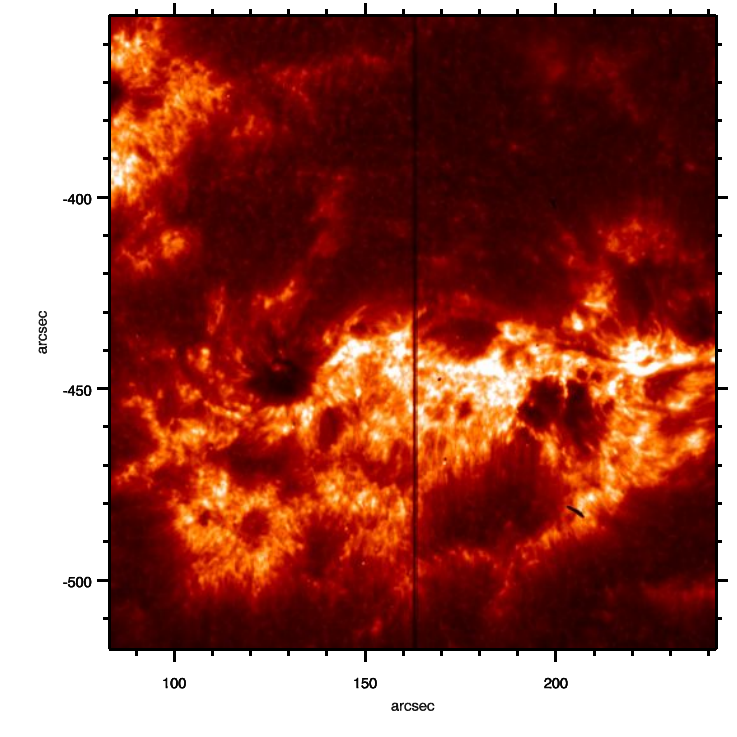}\includegraphics[width=0.5\textwidth,clip=]{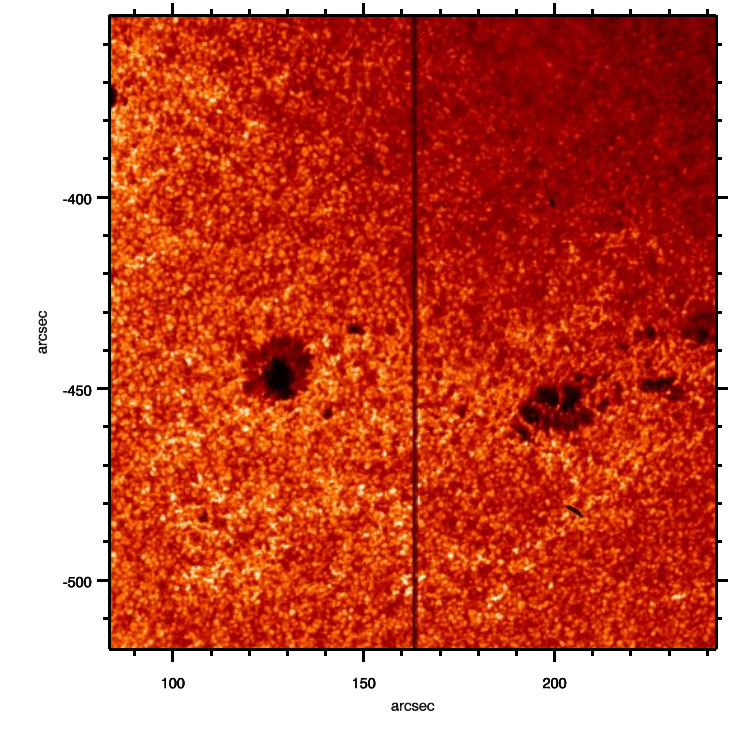}}
              \caption{IRIS SJI 2796\,\AA\ and 2830\,\AA\ images of NOAA
                AR 11817 taken on 14 Aug 2013 at 1850 UT. These images are senstive to plasma of the upper
              chromosphere (2796\,\AA) and upper photosphere (2830\,\AA). The
              upper chromospheric image also contains contributions
              from the upper photosphere to the middle chromosphere,
              which are dominant in more quiet regions. The
              dark vertical line in the middle of the images is the
              location of the slit.  Corresponding spectra are shown in
            Figures~\ref{fig1}, \ref{fig2}, \ref{fig3}.}
   \label{fig5}
   \end{figure}

\begin{figure}    
  \centerline{\includegraphics[width=1\textwidth,clip=]{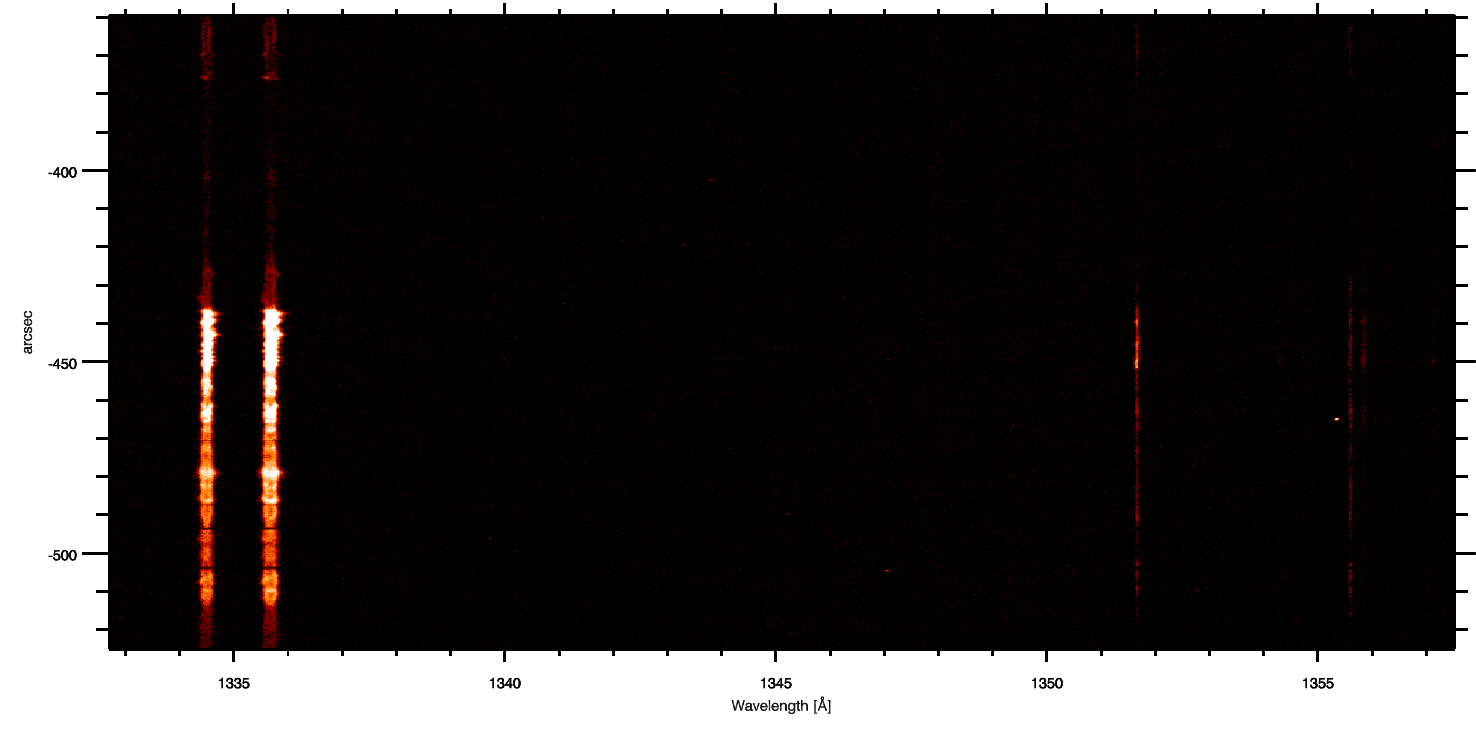}}
              \caption{IRIS FUV 1 spectrum of NOAA AR 11817 taken on 14 Aug 2013 at 1850
              UT. The two strong lines around 1334-1336\,\AA\ are C {\sc ii}
              lines that are formed in the upper chromosphere and low
            transition region. The lines longward of 1350\,\AA\ are Cl
            I, O {\sc i} and
            C {\sc i} lines that are formed in the low to middle
            chromosphere.  Corresponding images are shown in
            Figures~\ref{fig4}, \ref{fig5}.}
   \label{fig1}
   \end{figure}

\begin{figure}    
  \centerline{\includegraphics[width=1\textwidth,clip=]{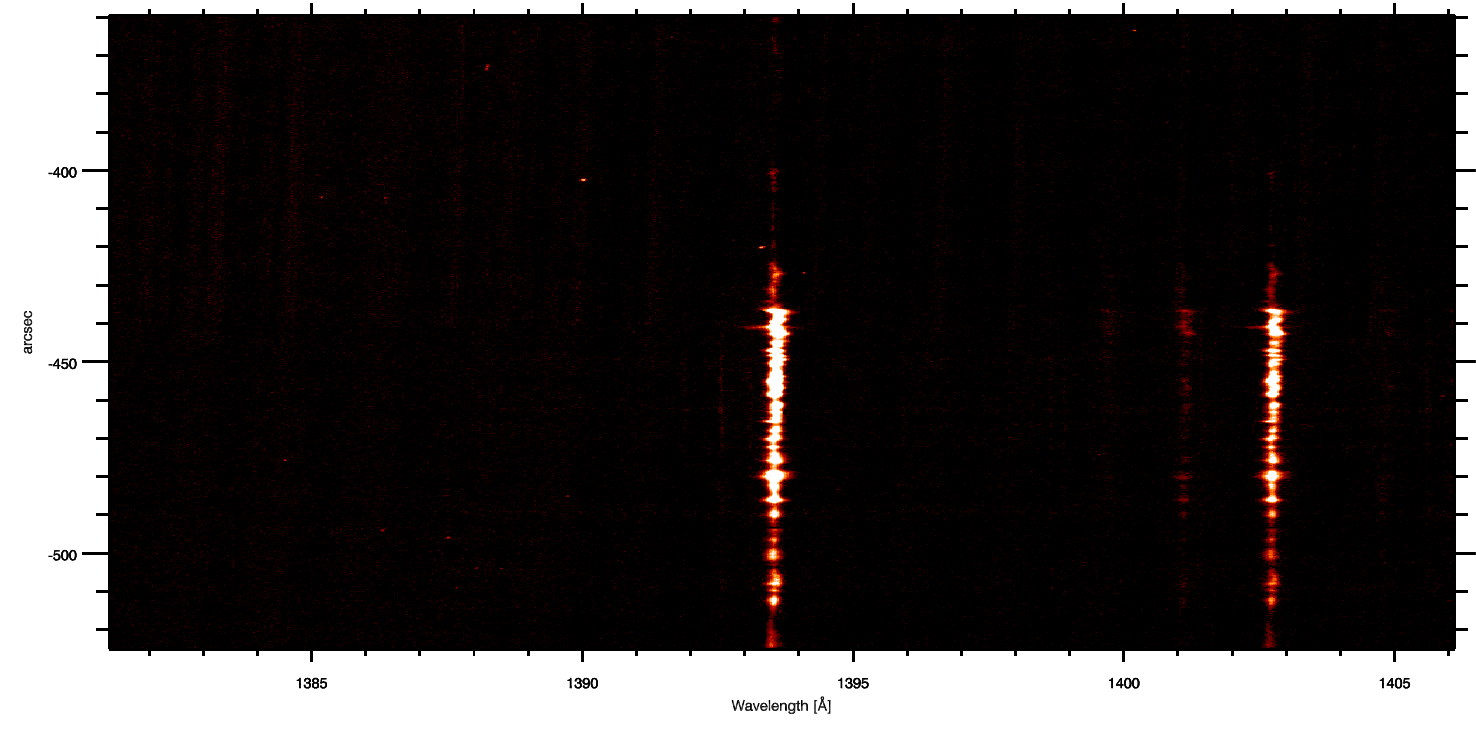}}
              \caption{IRIS FUV 2 spectrum of NOAA AR 11817 taken on 14 Aug 2013 at 1850
              UT. The strongest lines are Si {\sc iv} lines, formed in the
              transition region (65\,000 K). The weak lines around the
              Si {\sc iv} 1402\,\AA\ line are O {\sc iv} lines which are formed
              around 150\,000K under equilibrium conditions. The very faint
              vertical lines in the upper half of the detector are an
              example of the noise pattern caused by electronic
              interference during camera readout (see
              Section~\ref{S-CCD}). Corresponding images are shown in
            Figures~\ref{fig4}, \ref{fig5}.}
   \label{fig2}
   \end{figure}

\begin{figure}    
  \centerline{\includegraphics[width=1\textwidth,clip=]{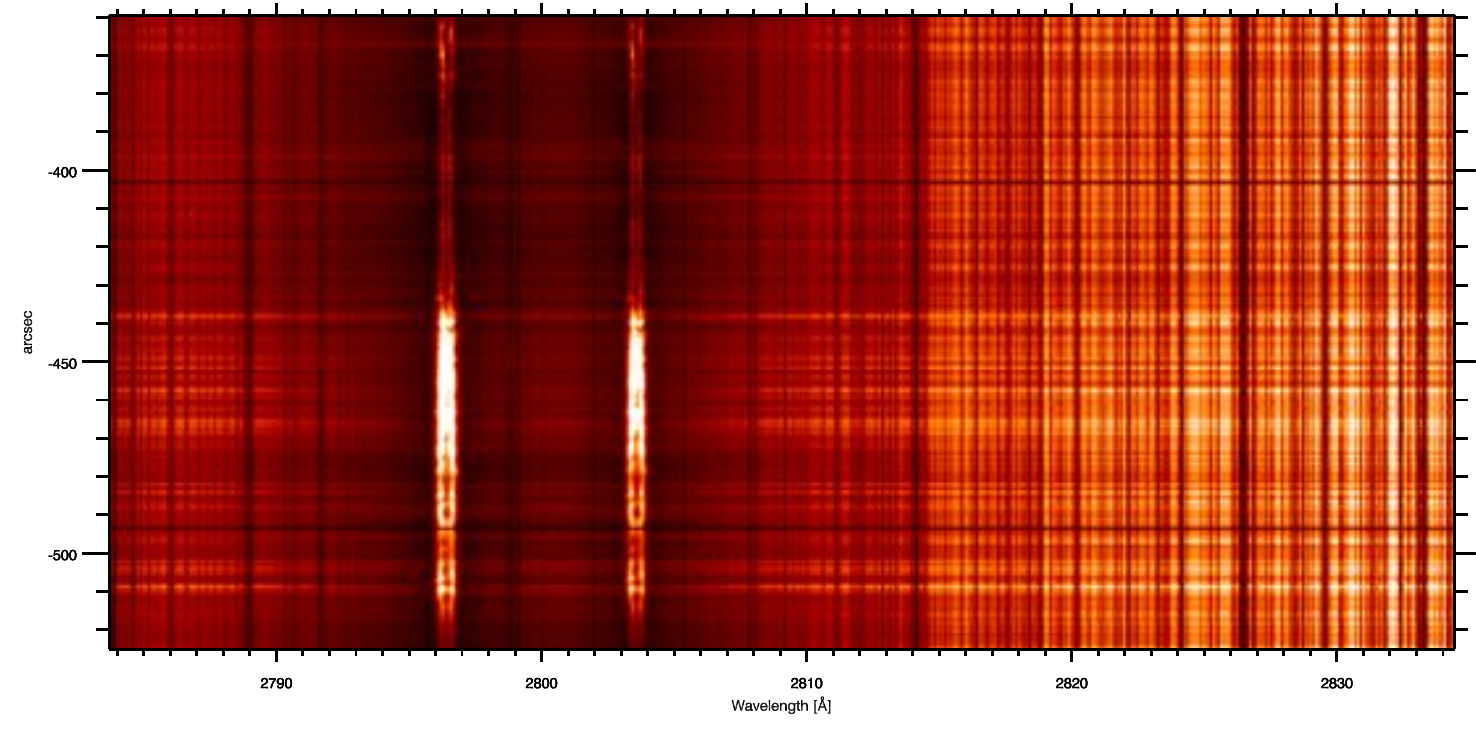}}
              \caption{IRIS NUV spectrum of NOAA AR 11817 taken on 14 Aug 2013 at 1850
              UT. The two strong emission lines are Mg {\sc ii} k 2796\,\AA\
              and Mg {\sc ii} h 2803\,\AA\, both formed over a range of
              heights from the upper photosphere to the upper
              chromosphere. This wavelength range also contains a multitude
              of photospheric lines. The thin horizontal lines are
              fiducial marks that allow for easy co-alignment. Corresponding images are shown in
            Figures~\ref{fig4}, \ref{fig5}.}
   \label{fig3}
   \end{figure}

\begin{figure}    
  \centerline{\includegraphics[width=1\textwidth,clip=]{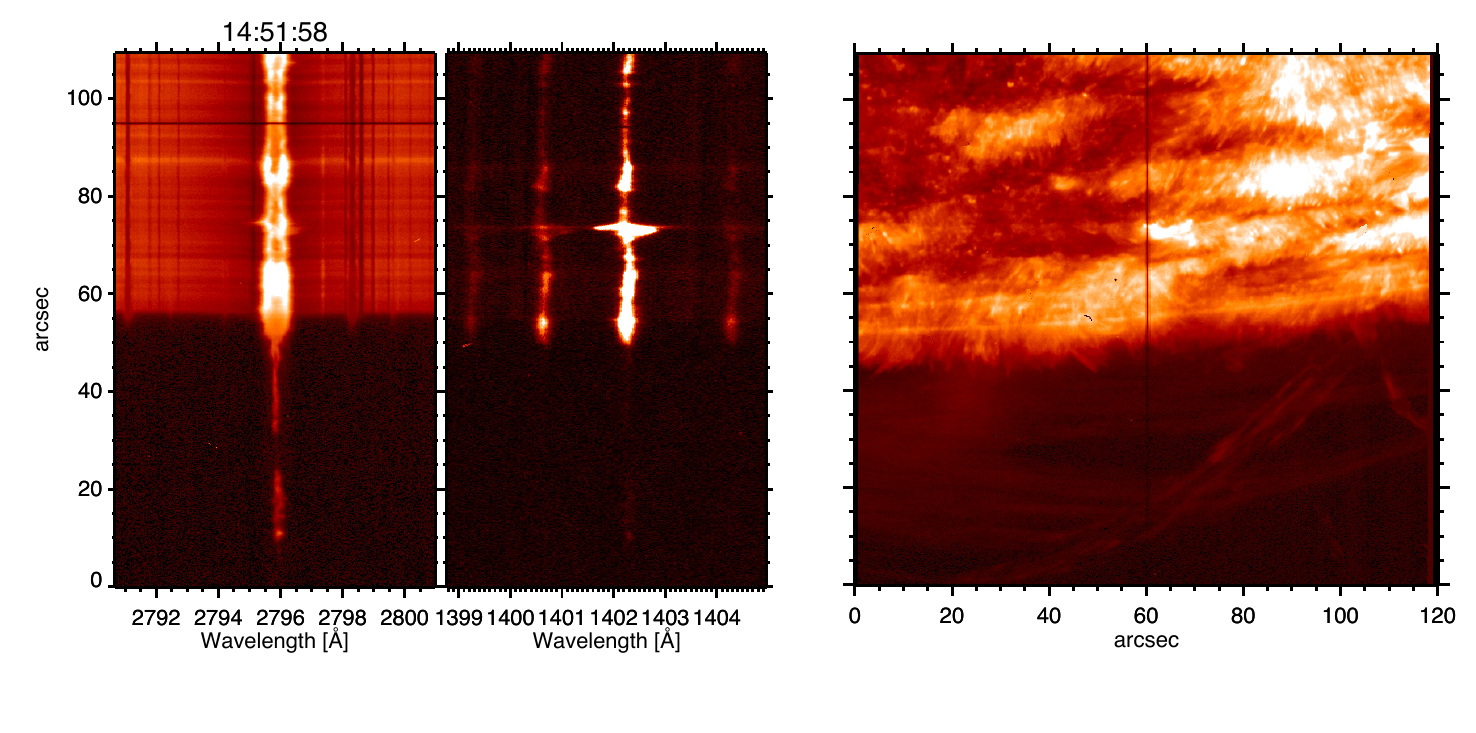}}
              \caption{IRIS spectra of Mg {\sc ii} k 2796\,\AA\ (left panel),
                Si {\sc iv} 1402\,\AA\ (middle panel), and 1400 SJI image
                (right panel) of an active region at the limb. The
                faint lines surrounding the bright Si {\sc iv} line in the
                middle panel are O {\sc iv} lines which can be used to
                determine densities in the transition region. Note the
              strong blue- and redward excursions in all spectral
              lines throughout the plage region. Off-limb the slit-jaw
              image captures spicules, coronal rain, and prominence
              material. The solar limb in the NUV spectrum (left
              panel) clearly shows the different range of formation
              heights for various spectral lines. The spacecraft was
              rolled by 90 degrees for these observations. The
              x and y coordinates are shown to indicate the scale of
              the image and spectra only.}
   \label{fig3b}
   \end{figure}

Despite the importance of the chromosphere and TR interface region for
solar activity, the heating of the corona, and the genesis of the
solar wind, this interface region remains poorly understood because 
it is highly complex and must be observed over a wide spectral range
(from the visible to the EUV). As a result, it
presents a challenging target for observers and modelers alike. The
transition between high and low plasma $\beta$
occurs somewhere between the photosphere and corona, so that in the
interface region, the magnetic field and plasma compete for dominance
(with a variety of impacts on, {\it e.g.}, waves, such as mode coupling,
refraction and reflection). Within this region, the density drops by six
orders of magnitude, and the temperature rapidly increases from 5000 to 1
million K, with strong gradients across the magnetic field evident
from high-resolution images and spectra of the chromosphere (see, {\it e.g.},
Figures~\ref{fig4}, \ref{fig5}, \ref{fig1}, \ref{fig2}, \ref{fig3},
and \ref{fig3b}).

The plasma transitions from partially ionized in the chromosphere to
fully ionized in the corona and shows evidence of supersonic and
super-Alfv{\'e}nic motions.  In addition, the chromosphere is partially
opaque, with non-local thermodynamic equilibrium (non-LTE) effects
dominating the radiative transfer, so that interpreting the radiation,
and determining the local energy balance and ionization state, is
non-intuitive and requires advanced computer models. The highly
dynamic nature of the chromosphere, as observed with Hinode
\cite{hinode} and
ground-based telescopes, further complicates attempts to better
understand the interface region. This is both because high cadence
observations are required (better than $\approx$ 20 seconds), and because the
ionization states of some elements ({\it e.g.}, hydrogen) react only slowly
to changes in the energy balance, and thus depend on the history of
the plasma. 

The launch of the IRIS mission opens a new window into the complex physics 
of the interface region. The spectral ranges that IRIS observes have
previously been studied at lower resolution using rockets
\cite{Bates69,Fredga69,Kohl76,Allen78,Morrill08,West11,Dere84}, balloons
\cite{Lemaire69,Samain85,Staath95}, or satellites
\cite{Doschek77,Bonnet78,Woodgate80,Lemaire73,Roussel82,Billings77,Poland83,Kingston82}. IRIS
draws on heritage solar instrumentation, such as the {\it Transition Region
and Coronal Explorer} (TRACE: \opencite{trace}), the {\it Helioseismic
and Magnetic Imager} (HMI: \opencite{hmi}) and the {\it Atmospheric Imaging
Assembly} (AIA: \opencite{aia}) onboard
the {\it Solar Dynamics Observatory} (SDO: \opencite{sdo}), and it exploits 
advances in novel, high-throughput, and high-resolution instrumentation, 
efficient numerical
simulation codes, and powerful, massively parallel supercomputers to
aid interpretation of the data.

In Section~\ref{S-over} we give an overview of the IRIS observatory and its
capabilities. We describe the science goals of the IRIS mission in Section~\ref{S-science}
and the instrument in Section~\ref{S-instrument}. The instrument observing approach and
the day-to-day operations are described in Sections~\ref{S-sequencer} and \ref{S-ops}, respectively. We discuss the preliminary
calibration of IRIS in Section~\ref{S-calibration}, while the data
processing is detailed in Section~\ref{S-data}. We finish with a description of the numerical-simulations approach
in the IRIS science investigation (Section~\ref{S-sims}). Our conclusions are given in Section~\ref{S-conclusions}. 
 
\section{IRIS Observatory} 
      \label{S-over}      

\subsection{IRIS Properties}
The conditions in the interface region present a significant
challenge to observers. Observations with {\it Hinode} and ground-based
instruments such as the {\it Interferometric Bidimensional
  Spectrometer} (IBIS: \opencite{ibis}) at the {\it Dunn Solar
  Telescope} (DST) and the {\it Crisp Imaging Spectropolarimeter}
(CRISP: \opencite{crisp}) at the {\it Swedish Solar Telescope} (SST: \opencite{sst}) indicate that
to properly capture the dynamics, fine-scale structuring, and
small-scale wave motions, observations
at a cadence better than 20 seconds and a spatial resolution of better
than 0.5~arcsec are required. To track the thermal evolution
of the heating and cooling plasma in the interface region, observations
need to simultaneously cover temperatures from the photosphere into
the corona. To diagnose plasma conditions such as velocity,
turbulence, non-thermal energy, or density, spectroscopic information
needs to be obtained at high resolution and high
signal-to-noise (S/N) ratios. Velocity fields must be determined to
an accuracy of 1 km~s$^{-1}$ ({\it e.g.}, through sub-pixel line centroiding), and to resolve non-thermal line broadening a
spectral resolution of order 5\,--\,10 km~s$^{-1}$ is required. An instrument with
these properties should be able to, {\it e.g.}, quantify the properties of
the waves that permeate this interface region \cite{DePontieu07,McIntosh11,Okamoto11,DePontieu12,McIntosh12a,Sekse13} and disentangle the
complexities of the multiple spectral components that have been
indirectly inferred using current spectroscopic data \cite{DePontieu09,Bryans10,Tian11,McIntosh12b}.

The observational requirements given above drove the IRIS mission 
to provide the following essential capabilities:
\begin{itemize}
  \item High spatial resolution (0.4~arcsec) spectroscopic and (context) imaging data
  over a field of view of at least 120~arcsec providing
  diagnostics from the photosphere to the corona, with a focus on the
  chromosphere and transition region.
\item A high signal-to-noise ratio for two second exposures for a
few select bright lines covering chromosphere and transition region
allowing velocity determination with 1 km~s$^{-1}$ accuracy and 3 km~s$^{-1}$
spectral pixels.
\item High cadence spectral (20~s) and imaging (10~s) observations covering a small region of the Sun
(5 $\times$ 120 arcsec) for periods of up to eight hours continuously.
\item Eclipse-free observations for up to eight months per year with about
  15 X-band passes per
day and an average data rate of 0.7 Mbit~s$^{-1}$.
\end{itemize}

These capabilities are met by a design that includes the following:
\begin{itemize}
\item A 19-cm Cassegrain telescope that feeds a dual-range UV
spectrograph (SG) and slit-jaw imager, with 0.16~arcsec pixels and four
$2061\times1056$ CCDs. 

\item A slit-jaw imager that includes four passbands with two
transition-region lines (C {\sc ii} 1335\,\AA\ and Si {\sc iv} 1400\,\AA),
one chromospheric line (Mg {\sc ii} k 2796\,\AA) and one photospheric
passband (2830\,\AA), covering a field-of-view of 175~arcsec $\times$ 175~arcsec.

\item A spectrograph with 0.33~arcsec wide and 175~arcsec long slit that covers FUV
passbands from 1332\,\AA\ to 1358\,\AA\ and 1389\,\AA\ to 1407\,\AA\ and an
NUV passband from 2783\,\AA\ to 2835\,\AA. These passbands include lines
formed over a wide range of temperatures from the photosphere (5000 K)
to the corona (1 to 10 million K).

\item CCD detectors with a full well of 150\,000 electrons, with a
camera readout noise of $<$20~e$^{-}$, and data compression that is nearly
lossless.

\item Instrument control software that allows for flexible rastering of the
slit across the Sun (up to 21 arcmin from disk center), onboard
summing, and various slit-jaw choices and cadences.

\item A baseline cadence of 3 seconds per spectral raster position, 5 seconds for
slit-jaw images.
\end{itemize}

These capabilities enable IRIS to observe the thermal
evolution of plasma from photospheric to coronal temperatures at the
spatio-temporal resolution required for the highly dynamic interface
region. The spectral, temporal, and spatial resolution, and spectral coverage and
effective areas of IRIS constitute significant advances over previous instrumentation.
 The IRIS throughput is more than an order of magnitude better than that
of previous spectrographs such as the {\it Solar Ultraviolet
  Measurements of Emitted Radiation} instrument (SUMER: \opencite{sumer}) or the {\it
  Extreme ultraviolet Imaging Spectrograph} (EIS: \opencite{eis}), both of which lack
slit-jaw imaging for context. In its typical operational mode, IRIS
obtains spectra about five to ten times faster than SUMER or EIS. The
effective spatial resolution of IRIS is 0.4~arsec, compared with the
Nyquist-limited resolution of two arcsec of EIS or SUMER. 
IRIS enables imaging of the interface region with ten
resolution elements for each of SDO/AIA's and 25 for
each of SOHO/SUMER or Hinode/EIS. The velocity resolution of IRIS is
more than three times better than SUMER and ten times better than EIS.

  \begin{figure}    
  \centerline{\includegraphics[width=1\textwidth,clip=]{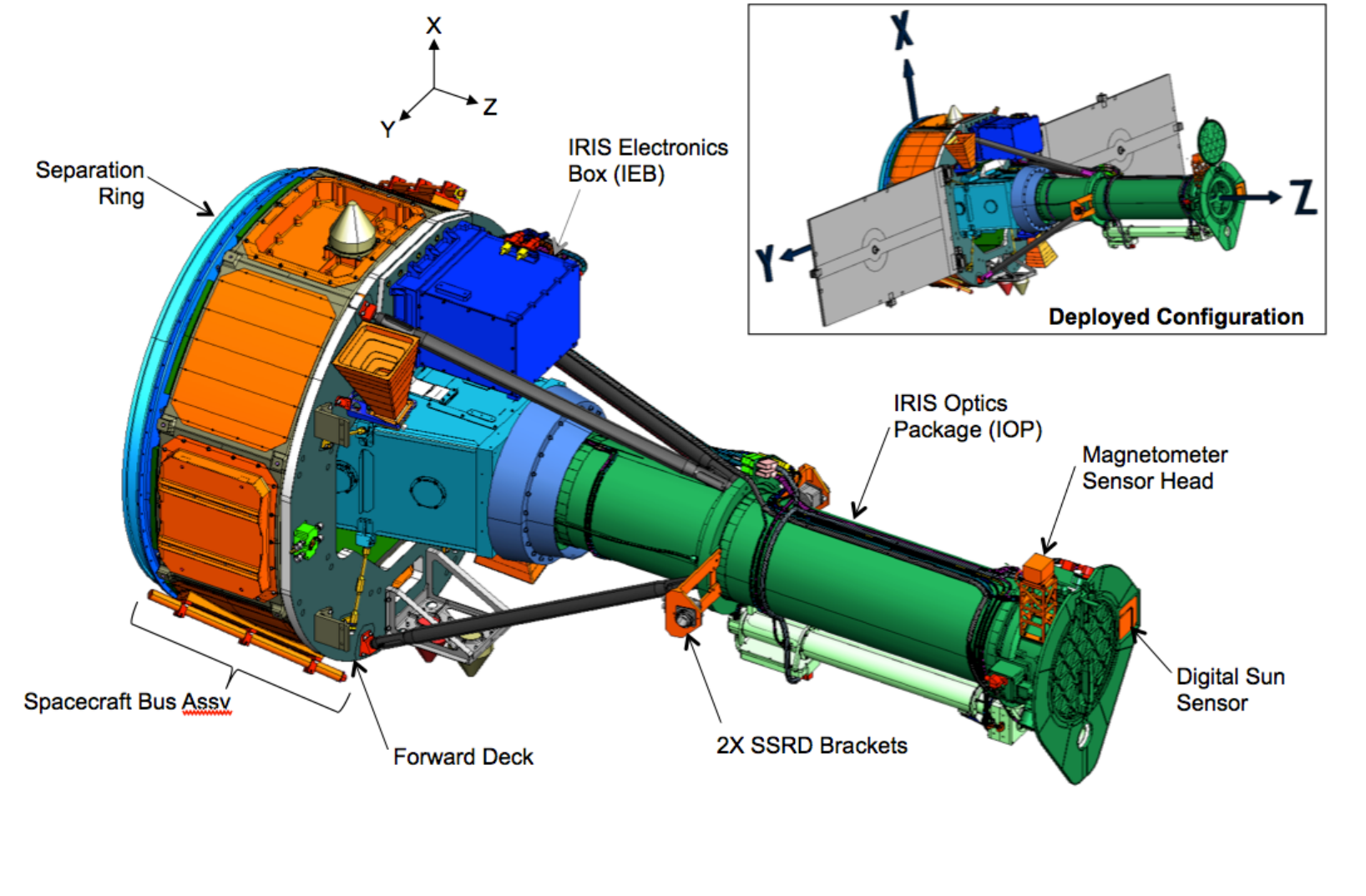}}
              \caption{Schematic view of IRIS showing the 19-cm UV
                telescope, with and without solar panels (for
                clarity). Light from the Cassegrain telescope (green)
                is fed into the spectrograph box (light blue). }
   \label{fig_obs}
   \end{figure}

\subsection{Orbit}
IRIS was launched by an Orbital Space Systems Pegasus-XL rocket on 27 June 2013
into a Sun-synchronous, low-Earth orbit with an
inclination of 97.9 degrees, perigee of 620 km and apogee of 670 km
with a 6 AM ascending node. For the first few years of operations, this orbit enables eclipse-free viewing
from the beginning of February until the end of October. The rest of
the year (November through January) the Earth blocks the IRIS view of
the Sun for part of the orbit. The baseline plan is that instrument
operations during eclipse season are not high priority, although power
and thermal conditions will be assessed during the first eclipse
season to determine the extent of IRIS operations during eclipse season. 

IRIS passes through the South Atlantic Anomaly (SAA)
and the northern and southern high latitude zones (HLZs). While SAA
passes lead to significantly increased numbers of energetic particle hits, the
IRIS detectors have been shielded sufficiently so that flying through
the HLZs has minimal impact on the quality of the data.


\subsection{Observatory}
The IRIS observatory mass is 183 kg with 87 kg for the instrument and
96 kg for the spacecraft bus. The IRIS spacecraft bus, designed and delivered
by Lockheed Martin Civil Space (formerly LM Sensing and Exploration Systems),
is a rigid design
whose frame is machined from a single piece of aluminum with honeycomb aluminum
forward and aft decks (Figure~\ref{fig_obs}).  Eight equipment bays house the spacecraft electronics 
boxes, reaction wheels, and the battery.  
The observatory measures about 2.18~m from the back of the
spacecraft bus to the front of the telescope. The solar arrays deployed
are 3.6~m tip to tip. The two solar arrays
measure 0.6 $\times$ 1.3~m each, with a total surface area of 1.7~m$^2$
producing 340~W. IRIS is three axis stabilized. The attitude control system 
(ACS) is gyroless using two star trackers, four reaction wheels, coarse and digital sun sensors, and a magnetometer.  The instrument
guide telescope provides a high-resolution pointing signal to the ACS during
normal science operations.  Magnetic torque rods are used to manage the 
momentum of the reaction wheels, transfering energy to the 
Earth's magnetic field as needed.  There is no propulsion system and there are
no consumables onboard.  The ACS can point the IRIS telescope boresight to 
any location on the solar disk or above the limb within 21 arcminutes of 
disk center, and roll the 
spacecraft (and thus, the 
spectrograph slit), up to $\pm90^\circ$ (at $0^\circ$ the slit is oriented 
parallel to N\,--\,S on the Sun). 
IRIS is equipped with two omnidirectional S-band
antennas for uplinking of commands and downlinking of engineering data, and an X-band antenna for
downlinking of science data. The S-band provides uplink at 2~kbit~s$^{-1}$
and downlink at 256~kbit~s$^{-1}$, whereas the X-band provides downlink at 15
Mbit~s$^{-1}$ including the overhead of Low Density Parity Checking (LDPC) 7/8s 
encoding. The effective downlink
rate is 13 Mbit~s$^{-1}$ (excluding overhead) during up to 15 passes per day
with the antennas of Kongsberg Satellite Services (KSAT) in Svalbard,
Norway, as well as some passes from NASA's Near Earth Network (NEN) in Alaska and Wallops.

\section{Science Overview} 
      \label{S-science}      

The IRIS investigation covers a broad range of scientific objectives
that focus on three major questions which form the foundation of the
IRIS investigation and have driven the requirements for the IRIS
instrument design. Below we list an overview of the science issues covered by
these three science questions.

\subsection{Which Types of Non-Thermal Energy Dominate in the
 Chromosphere and Beyond?}

We still do not know which modes of non-thermal energy power the
chromosphere, TR, corona, and solar wind. We know that waves,
electrical curents, and magnetic reconnection all may release
substantial energy, and that non-thermal particles, resistive
dissipation, and wave damping occur. Yet, it remains unclear how much
each of these contributes, how that depends on local conditions, and
how and where the conversion of non-thermal to thermal energy happens in detail.

\subsubsection{Waves}

A variety of different wave modes have been observed in the past few
years, from mostly acoustic modes and atmospheric gravity waves to
Alfv\'en waves. However, there are still many unresolved questions
regarding the role of these waves in the solar atmosphere. 

For example, the role of acoustic power in the heating of the chromosphere and its
potential role in the corona remains a puzzle, with some researchers reporting too
little power and others arguing there should be enough. Most of the
power in these modes lies in the range of the five minute $p$-mode spectrum
and the 10\,--\,15~minute time scale of granular evolution, but waves have
been measured for periods as short as $\approx$25~seconds with some observations
suggesting that these high frequency waves carry a large, potentially
dominant, energy flux. In addition, the normally evanescent $p$-mode
oscillations may play a major role in the energy balance of the
chromosphere in and around magnetic-field concentrations 
\cite[so-called “acoustic portals”]{Jefferies2006}. The propagation of these
sound waves along magnetic-field lines leads to the formation of
slow-mode magnetoacoustic shocks that propel plasma upwards to form
so-called dynamic fibrils \cite{Hansteen2006,DePontieu2007,Rouppe2007,Kato2011}. The role of these
shocks in heating the magnetic chromosphere remains unknown but may
be quite significant, based on analysis of spectra of the Ca {\sc ii} H line \cite{Beck2008}. Similarly, the enormous energy flux carried by
atmospheric gravity waves in the internetwork \cite{Straus2008}
and the potential for mode coupling to other wave modes ({\it e.g.}, Alfv\'en
waves) around magnetic-field concentrations \cite{Cally2008}
highlights another potential source of non-radiative energy.

Various types of mostly transverse MHD waves (such as Alfv\'en waves)
are likely more efficient at traveling into the corona and solar
wind, but their detailed energy budget remains mysterious. For example,
relatively low-frequency ($\approx1$ to 8 mHz) transverse and torsional
(Alfv\'enic) motions have now been measured in the chromosphere and
corona \cite{Tomczyk07,DePontieu07,Tomczyk09,McIntosh11,Liu11,DePontieu12,Wedemeyer12,Su13,Verwichte13},
but it is unclear what role higher frequency waves play, how these
waves are generated and how they are dissipated.


IRIS spectroscopic measurements of waves as they propagate from the photosphere through
the chromosphere and transition region into the corona will establish an
energy flux budget, allowing us to disentangle wave mode coupling,
dissipation and propagation. 

\subsubsection{Currents and Reconnection}

Establishing the energy associated with resistive dissipation of
electrical currents within the chromosphere has been
challenging. These currents are expected to be intense in compact
fibrils. Any dissipation within the chromosphere can be expected to
be associated with rapid temperature increases. Clearly, tracing the
thermal evolution of chromospheric plasma holds large potential for
providing information on how much electromagnetic energy is
transformed into heat and how much energy propagates downwards from
the corona as thermal energy or energetic particles in events ranging
from microflares to X-class flares. Both of these types of processes
are expected to occur, but we do not know their relative roles under a
variety of conditions.

The multi-fluid aspects introduced by the partial ionization of the
chromosphere can dramatically alter the way in which electrical
currents (and magnetic fields) are generated and dissipated (see,
{\it e.g.}, \opencite{Arber07}) and thus provide a promising method of
providing energy to the plasma. Various examples have recently been
studied. For example, \inlinecite{Goodman10} focused on the
dissipation of short period shock waves from ion-neutral resistivity,
whereas \inlinecite{Fontenla08} studied the Farley\,--\,Bunemann instability
(but see \opencite{Gogoberidze09}) which depends critically on the ion
magnetization (and thus the ion\,--\,neutral collision frequency and
ionization degree). 
The importance of ion\,--\,neutral coupling has also been illustrated in 
advanced numerical models of the coupled solar atmosphere in which the
chromosphere is self-consistently heated by ambipolar diffusion of
currents generated through braiding of the magnetic field from
magneto-convection \cite{Martinez12}.

A major unresolved issue is the role of reconnection of strong
network or plage magnetic fields with the
ubiquitous weak magnetic fields of mixed polarity on granular scales
\cite{Lites08} in energizing the low solar atmosphere. Because the
amount of weak flux that emerges on granular scales per day is much
larger than the flux associated with active region emergence over a
full solar cycle, it is clear that if even a fraction of the weak
fields reconnect with strong network or plage fields, the resulting
energy release could be a dominant contributor to chromospheric or
even coronal heating. Yet we do not know how much of this flux is
dissipated in the solar atmosphere before most of it is recycled below
the surface of the Sun. 

IRIS measurements of chromospheric and coronal heating will help
determine the sites of enhanced dissipation. Coordinated magnetic
field observations from ground and space-based telescopes will reveal
the role of magnetic field dissipation in the heating of the
chromosphere and corona.

\subsection{How Does the Chromosphere Regulate the Mass and Energy Supply
  to the Corona and Heliosphere?}

The heating of the solar corona and the acceleration of the solar wind
remain mysterious. IRIS contributes to our understanding by showing us
in what form non-radiative energy is transmitted by the chromosphere
into the corona. The studies described above help constrain wave energy
fluxes, while imaging of the chromosphere and photosphere combined
with modeling constrains the forcing of the field into braids (see,
{\it e.g.}, \opencite{Cirtain13}, \opencite{Guerreiro13}), and
twists by flows. IRIS also sheds light on how and where non-thermal
energy is first released. Non-radiative energy deposited into the
corona is lost both by radiation and transfer into the lower layers of
the solar atmosphere. The transport to the lower layers can be thermal
conduction, in which a relatively smooth temporal evolution can be
expected, leading to evaporation of warming material from the top of
the chromosphere. Alternatively, energy can propagate downward as
energetic particles that can penetrate into the high chromosphere:
such particle precipitations are likely more intermittent as they
rapidly shift in position. Recent high-resolution observations in the transition region
show direct evidence of this type of process (under some magnetic field
conditions) in active regions \cite{Testa13,Winebarger13}. 

Other recent studies suggest that heating to
coronal temperatures may occur primarily in the chromosphere and
transition region, as evidenced, {\it e.g.}, by chromospheric jets such as
spicules \cite{DePontieu11}, and their faint coronal counterparts, visible only as slight
enhancements in the blue wing of coronal spectral lines \cite{DePontieu09}. 

IRIS contributes to all of these issues by providing high resolution
and high signal-to-noise spectra and context images of spectral lines formed over a wide range
of temperatures. Such measurements help reveal complex line shapes
that arise along the line-of-sight in an interface region in which
strong explosive upflows, gentle evaporative flows and downflows all
co-exist \cite{McIntosh12b}.

IRIS high-cadence observations significantly improve our ability to
determine how much wave power is available to heat and accelerate the
plasma in open-field regions at the base of the solar wind.
Measurements of transverse and longitudinal waves, and their possible
couplings to the solar wind, also provide constraints to models of the
origin of turbulence in the heliosphere \cite{Axford1999,Tu1997,Cranmer2007}.

  \begin{table}
   \caption{IRIS instrument characteristics}
   \label{table_instrument}
    \begin{tabular}{ll}     
      \hline                   
      Primary diameter & 19 cm\\
    Effective focal length & 6.895 m\\

      Field of view & 175 $\times$ 175 arcsec$^2$ (SJI) \\
     & 0.33 $\times$ 175 arcsec$^2$ (SG -- slit)\\
     & 130 $\times$ 175 arcsec$^2$ (SG -- raster)\\

      Spatial scale (pixel) & 0.167 arsec \\

      Spatial resolution & 0.33 arcsec (FUV) \\
      & 0.4 arcsec (NUV) \\

      Spectral scale (pixel) & 12.8 m\AA\ (FUV) \\
      & 25.6 m\AA\ (NUV) \\

      Spectral resolution & 26 m\AA\ (FUV SG) \\
      & 53 m\AA\ (NUV SG) \\
      Bandwidth & 55 \AA\ (FUV SJI) \\
      & 4 \AA\ (NUV SJI) \\
    
      CCD detectors & Four e2v 2061 $\times$ 1056 pixels, thinned, back-illuminated \\
      CCD cameras & Two 4-port readout cameras (SDO flight spares)\\

      Detector full well & 150\,000 electrons \\

      Typical exposure times & 0.5 to 30 seconds \\

      Flight Computer & BAe RAD 6000\\

      Mass&\\
      \hspace{0.5cm} Instrument & 87 kg\\
      \hspace{0.5cm} Spacecraft & 96 kg\\
      \hspace{0.5cm} Total      & 183 kg\\

     Power & \\
      \hspace{0.5cm} Instrument & 55 W\\
      \hspace{0.5cm} Spacecraft & 247 W\\
      \hspace{0.5cm} Total      & 302 W\\     

     Science Telemetry & \\
     \hspace{0.5cm} Average downlink rate & 0.7 Mbit~s$^{-1}$\\
     \hspace{0.5cm} X-band downlink rate & 13 Mbit~s$^{-1}$\\
     \hspace{0.5cm} Total data volume   & $\approx$ 20 Gbytes
     (uncompressed) per day\\     
     \hline
    \end{tabular}
   \end{table}

\subsection{How Does Magnetic Flux and Matter Rise Through the Lower
  Atmosphere, and What Role Does Flux Emergence Play in Flares and
  Mass Ejections?}

As magnetic field breaches the surface of the Sun, it begins to
interact and reconnect with the existing field into which it
intrudes. 
IRIS observations of flows help reveal the 3D flow
patterns associated with flux emergence. The upflows of a few km~s$^{-1}$ in
the low chromosphere, the associated downdrafts of the draining
archfilament system at larger speeds, and the field expansion in the
chromosphere and corona at Alfv\'en speeds are all measurable with
IRIS. These observations will also help to uncover the reconnection processes that
happen early in the life of active regions down to small
granular-scale fields.

Emerging flux appears to play a key role in many flare
phenomena. Observing flux as it breaches the surface and the lower
chromosphere is an important diagnostic to understand flare
initiation: chromospheric reconnection has been inferred in cases
where emerging flux triggers a flare, and particle precipitation
results in chromospheric ribbons both as tracers of coronal
reconnections and as power sources for chromospheric
evaporation. Observations of such flux emergence with Hinode suggest
that the site of first energy release in a major flare occurred where
strong electrical currents emerged. The thermal coverage and high
resolution of IRIS allow insight into how the atmosphere evolves in
the case of
such a rapid energy release, and help constrain why flares are
initiated where they are.


Spectroscopic observations of solar eruptions such as jets, flares, and CMEs can provide valuable information on the dynamics and plasma properties of the erupted materials \cite{Harra2007,Tian2012,Young2013}. IRIS' higher spectral resolution, higher cadence, and chromospheric and transition region response make it a unique instrument to study the initiation and early evolution of solar eruptions. 

\section{Instrument Overview} 
      \label{S-instrument} 

Portions of the IRIS instrument are described in some detail in
pre-launch articles by \inlinecite{Wuelser12} and \inlinecite{Podgorski12} and
key parameters are provided in Table \ref{table_instrument}. Here we
briefly summarize the instrument capabilities and refer the reader desiring
more technical details to the above mentioned spectrograph and telescope articles.

  \begin{table}
   \caption{IRIS spectrograph channels. Dispersion, Camera Electronics
     Box (CEB), and Effective Area (EA) vary for the three band passes.}
   \label{table_sg}
    \begin{tabular}{lllllllll}     
      \hline                   
      Band & Wavelength&Disp.
      &FOV&Pixel&CEB&Shutter&EA&Temp.
      \\
      & [\AA] & [m\AA & [$^{\prime\prime}$] & [$^{\prime\prime}$] & & &[cm$^2$]&[log{\it T}]\\
      &      & pix$^{-1}$]& & & & &\\
\hline
FUV 1 & 1331.7--1358.4&12.98&175&0.1663&1& FUV SG&1.6&3.7--7.0\\
FUV 2 & 1389.0--1407.0&12.72&175&0.1663&1& FUV SG&2.2&3.7--5.2\\
NUV &   2782.7--2835.1&25.46&175&0.1664&2& NUV SG&0.2&3.7--4.2\\
   \hline
    \end{tabular}
   \end{table}

\begin{figure}    
  \centerline{\includegraphics[width=1\textwidth,clip=]{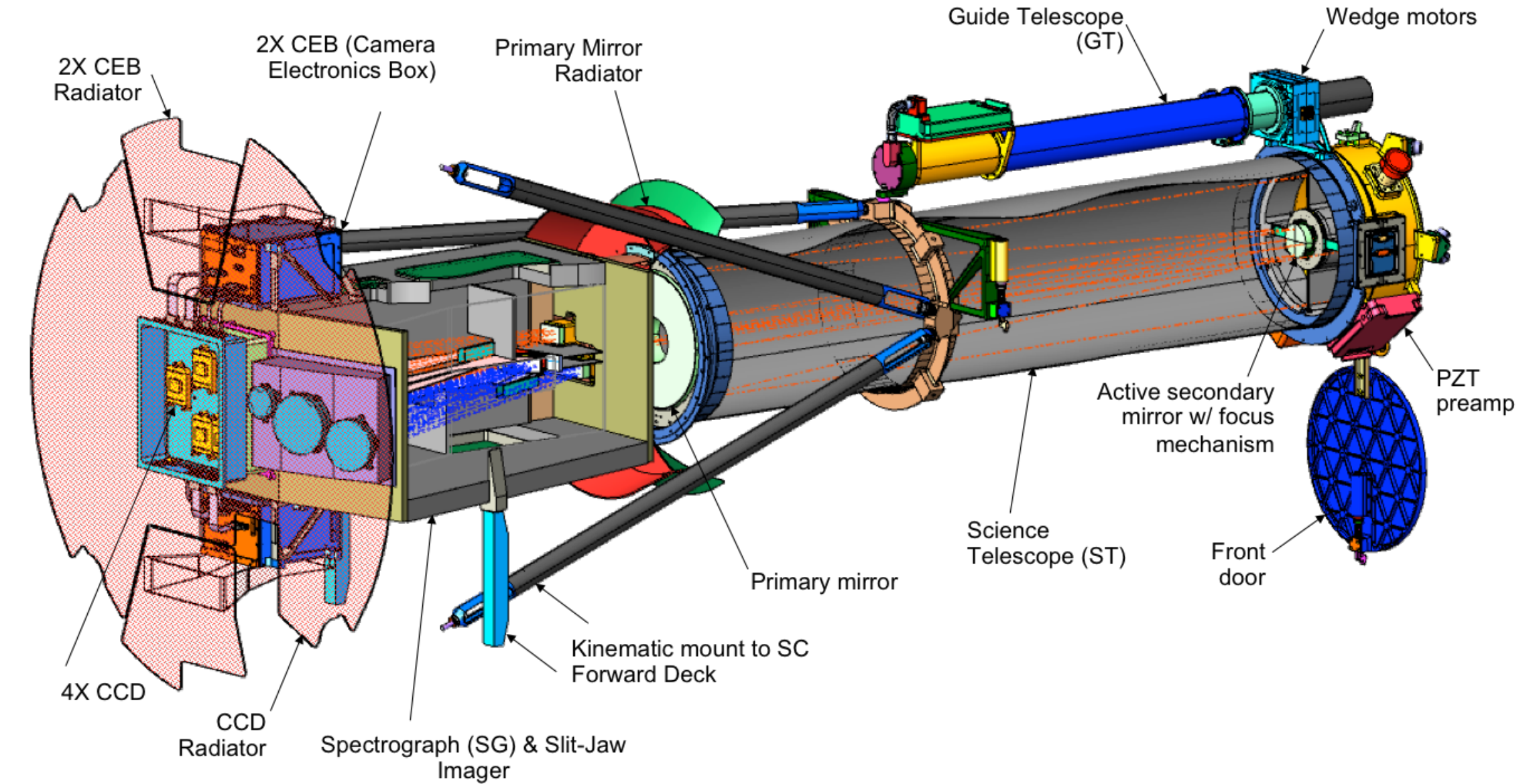}}
              \caption{Conceptual design of the IRIS instrument.
                Sunlight enters from the right.  For the flight
                design the telescope and guide telescope assemblies
                are rotated 180 degrees about the instrument axis
                relative to the spectrograph box (see
                Figure~\ref{fig_obs}).}
   \label{fig_telescope}
   \end{figure}

\begin{figure}    
  \centerline{\includegraphics[width=1\textwidth,clip=]{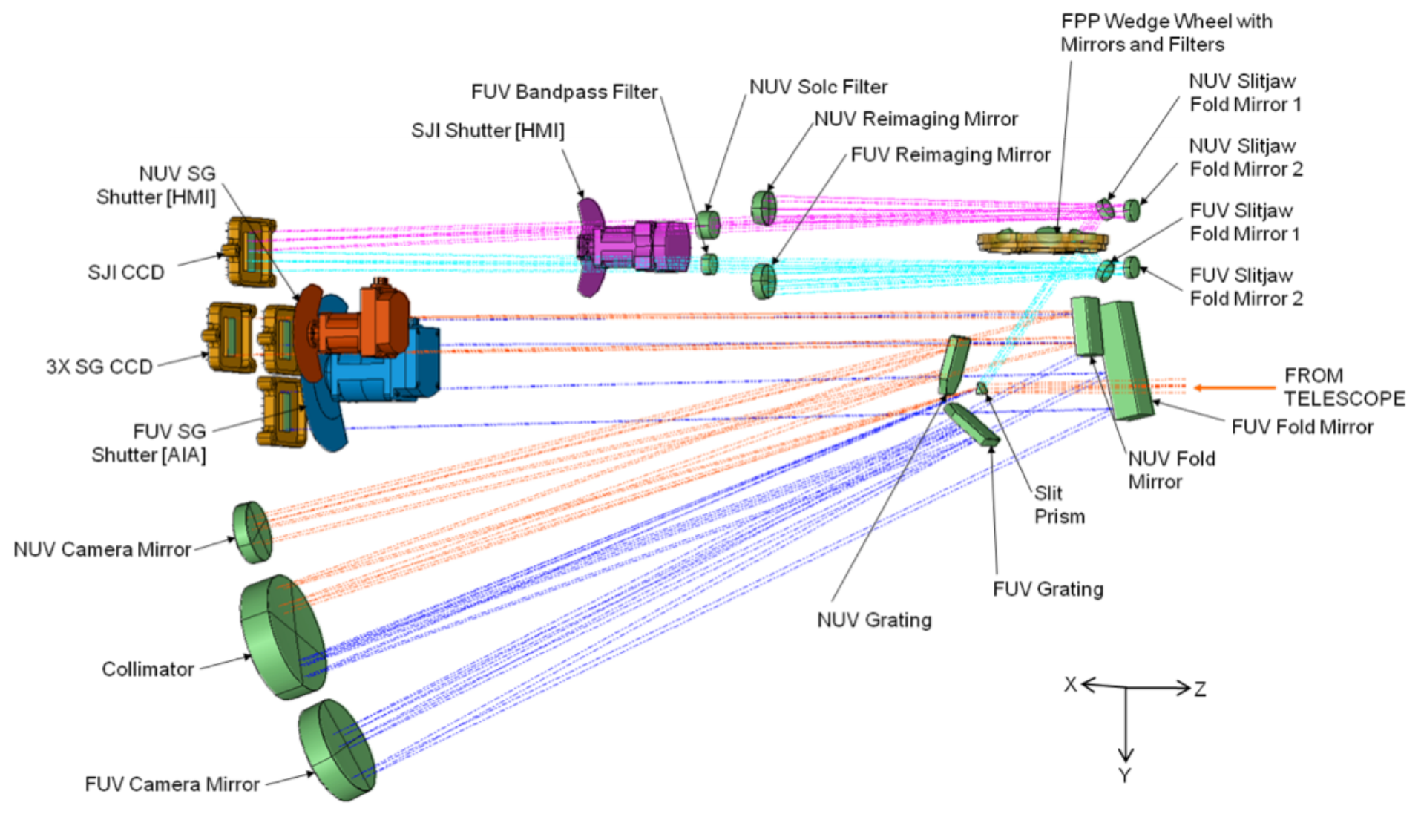}}
              \caption{Path taken by
                light in the FUV spectrograph (dark blue), NUV
                spectrograph (orange), FUV slitjaw (light blue) and
                NUV slitjaw (purple) path.}
   \label{fig_spectrograph}
   \end{figure}

The IRIS instrument uses a Cassegrain telescope with a 19-cm
primary mirror and an active secondary mirror with a focus mechanism 
(Figures~\ref{fig_obs}, \ref{fig_telescope}). The telescope has a
field of view of about 3 arcmin $\times$ 3 arcmin and 
feeds far UV (FUV, from 1332 to 1407\,\AA) and near UV (NUV, from 2783
to 2835\,\AA) light into a spectrograph box
(Figures~\ref{fig1}, \ref{fig2}, \ref{fig3}, \ref{fig3b}). Dielectric coatings throughout the optical path ensure visible and IR
radiation is suppressed. Most of the solar energy passes through the
ULE substrate of the primary mirror and is radiated back into space. The
FUV and NUV light follow several paths in the spectrograph box, as
illustrated in Figure ~\ref{fig_spectrograph}:

\begin{itemize}

\item Spectrograph (SG): light passes through a slit that is 0.33 arcsec
  wide and 175 arcsec long, and is dispersed onto either an NUV or an FUV grating.  Light from the FUV grating is collected by two CCDs and light from the NUV grating is collected by a separate CCD (Table \ref{table_sg}).

\item Slit-jaw Imager (SJI): light is reflected off the reflective area around
  the slit (“slit-jaw”), passing through or reflecting off broadband
  filters in the filterwheel onto a fourth CCD to produce an image of the
  scene around the slit in six different filters (two for ground testing, four for
  solar images, Table \ref{table_sji1}).
\end{itemize}

  \begin{table}
   \caption{IRIS slitjaw channels. Filterwheel positions can be either
   transmitting (T) or reflecting/mirrors (M).}
   \label{table_sji1}
    \begin{tabular}{lllllllll}     
      \hline                   
      Band-&FW&Name&Center&Width&FOV&Pix.&EA&Temp.
      \\
      pass& &
      &[\AA]&[\AA]&[$^{\prime\prime}\times^{\prime\prime}$]&[$^{\prime\prime}$]&[cm$^2$]&[log
      {\it T}]\\
\hline
Glass & 1 T       &5000 &5000&broad&175$^2$&0.1679&-&-\\ 
C {\sc ii}  & 31 M      &1330 &1340&55&175$^2$   &0.1656&0.5&3.7--7.0\\ 
Mg {\sc ii} h/k& 61 T   &2796 &2796&\ 4&175$^2$   &0.1679&0.005&3.7--4.2\\ 
Si {\sc iv} & 91 M      &1400 &1390&55&175$^2$   &0.1656&0.6&3.7--5.2\\ 
Mg {\sc ii} wing & 121 T&2832 &2830&\ 4&175$^2$   &0.1679&0.004&3.7--3.8\\ 
Broad & 151 M     &1600W&1370&90&175$^2$   &0.1656&-&-\\
\hline
    \end{tabular}
   \end{table}

As indicated above, in the spectrograph focal plane there are four
back-thinned CCD sensors with $2061\times1056$ pixels, three in the
spectrograph path (two for FUV, one for NUV), and one CCD at the
slit-jaw focal plane.  Each 13-$\mu$m pixel corresponds to 0.167 arcsec
in the spatial direction, and 12.8\,m\AA\ in the spectral direction for
FUV spectra, and 25.5\,m\AA\ for NUV spectra (the detailed plate scales
for each channel are given in Tables \ref{table_sg} and \ref{table_sji1}).
IRIS has an effective
spatial resolution between 0.33 (FUV) and 0.4 arcsec (NUV), and an
effective spectral resolution of 26\,m\AA\ in the FUV and 53\,m\AA\ in the
NUV.  The four CCDs are controlled by two camera electronics boxes
(CEBs). Exposure times are controlled by three different mechanical shutters
(FUV, NUV and SJI). The IRIS telescopes also includes an active secondary 
mirror (driven by piezoelectric transducers or PZTs) for fine-scale pointing
and image stabilization. The active secondary is pointed in response to
signals from a guide telescope, which is mounted on the bottom of the
telescope (Figure~\ref{fig_obs}).

The field of view imaged by IRIS is $175\times175$ arcsec$^2$ for the
slit-jaw images. To produce a raster of spectra of the Sun, the IRIS
active secondary mirror is scanned (using PZTs) in the direction
perpendicular to the slit, causing different regions of the Sun to be
exposed onto the slit. The slit scan range is $\pm65$ arcsec, so that
the maximum field of view of the IRIS rasters is $130\times175$ arcsec$^2$
for the SG. The IRIS slit is nominally oriented parallel to the solar
rotation axis (solar North), but the spacecraft can be rolled to any
angle between $-90^\circ$ and $+90^\circ$ from solar North for extended
periods of time. 

Many parts of the IRIS instrument are based on heritage from TRACE, AIA,
and HMI as noted below. Extensive use of designs and parts from these
previous missions reduced the need for lengthy life testing (for
mechanisms) and enabled the modest schedule and budget of a small
explorer to be achieved (45 months from start of Phase B until launch).

\subsection{Telescope}

The IRIS telescope is a 19-cm Cassegrain telescope with an active
secondary mirror. It is based on the AIA telescope design, but has a
longer focal length and uses a different thermal approach
\cite{Park12,Hertz12}.  Unlike the AIA design, sunlight enters the
IRIS telescope unattenuated.  At the primary mirror, most of the visible
and infrared light passes through the trasparent ULE mirror substrate and
is collected by a heat sink.  The front of the primary mirror has a
dielectric coating that reflects UV light to the secondary mirror 
which is subsequently focused on the spectrograh slit.  The telescope
is described in more detail by \inlinecite{Podgorski12}.

\subsection{Spectrograph}

The telescope feeds light into the spectrograph box
(Figure~\ref{fig_telescope}) which contains the Czerny--Turner spectrograph
(Figure~\ref{fig_spectrograph}). The light from the telescope is focused
on the slit assembly.  The slit assembly is a prism that has a
reflective coating, which also contains the slit.  The reflective coating
directs the light into the slit--jaw imager path.  Light that goes
through the slit into the prism is dispersed, directing FUV light in the
1332\,--\,1407\,\AA\ range and NUV light in the 2783\,--\,2835\,\AA\ wavelength
range onto separate parts of the collimator mirror. The
slit/predisperser prism assembly ensures that both FUV and NUV passbands
image the same region on the Sun within the 1/3 $\times$ 175 arcsec$^2$
entrance slit.

After the collimator, the FUV and NUV SG spectrograph beams are fed to separate gratings, 
camera mirrors, and detectors (Table~\ref{table_sg} and
Figure~\ref{fig_spectrograph}). The FUV and NUV gratings, fabricated by
Horiba Jobin--Yvon, have a groove density of 3600 lines~mm$^{-1}$ and are described 
in more detail by
\inlinecite{Wuelser12}. The FUV and NUV SG beams have separate shutters
and are recorded onto three separate CCDs -- two for the FUV and one for the 
NUV. The two FUV CCDs observe two separate wavelength ranges: one that
includes two bright C {\sc ii} lines (1332\,--\,1358\,\AA), and another that
contains Si {\sc iv} and O {\sc iv} lines (1389\,--\,1407\,\AA). These two FUV CCDs are
controlled by the same CEB and read out as if they were one CCD. The NUV passband from 2783\,--\,2835\,\AA\ is
recorded by a CCD that is controlled by a different CEB (which also
reads out the SJI CCD).

  \begin{table}
   \caption{Thermal coverage of IRIS spectrograph}
   \label{table_lines}
    \begin{tabular}{llllll}     
      \hline                   
      Ion&Wavelength&Dispersion&Log{\it T}&Passband&CEB\\
         &[\AA]     &[m\AA    &[log K]&  &\\
         &          &pix$^{-1}$]& & & \\
\hline
Mg {\sc ii} wing&2820  &25.46&3.7-3.9&NUV & 2\\
O {\sc i}       &1355.6&12.98&3.8&FUV 1& 1\\
Mg {\sc ii} h   &2803.5&25.46&4.0&NUV & 2\\
Mg {\sc ii} k   &2796.4&25.46&4.0&NUV & 2\\
C {\sc ii}      &1334.5&12.98&4.3&FUV 1& 1\\
C {\sc ii}      &1335.7&12.98&4.3&FUV 1& 1\\
Si {\sc iv}     &1402.8&12.72&4.8&FUV 2& 1\\
Si {\sc iv}     &1393.8&12.72&4.8&FUV 2& 1\\
O {\sc iv}      &1399.8&12.72&5.2&FUV 2& 1\\
O {\sc iv}      &1401.2&12.72&5.2&FUV 2& 1\\
Fe {\sc xii}    &1349.4&12.98&6.2&FUV 1& 1\\
Fe {\sc xxi}    &1354.1&12.98&7.0&FUV 1& 1\\
\hline
    \end{tabular}
   \end{table}

IRIS spectral rasters are formed by scanning across the solar surface
using the PZTs to change the orientation of the secondary
mirror. Typical observing programs include both the FUV and NUV
SG passband. The FUV and NUV spectral bandpasses cover spectral lines and
continua that in the solar atmosphere are formed over a range of temperatures
log {\it T} [K] = 3.7\,--\,7. Table~\ref{table_lines} describes these
lines in more detail. The brightest lines in the SG are the C {\sc ii} lines around
1335\,\AA, Si {\sc iv} 1394\,\AA, Si {\sc iv} 1403\,\AA, Mg {\sc ii} k 2796\,\AA, and Mg {\sc ii} h
2803\,\AA. These are the lines that are included in routine,
high-cadence raster scans where exposure times are of the order of two
seconds. The O {\sc iv}, Fe {\sc xii}, and Fe {\sc xxi} lines are fainter and require longer
exposure times.

\subsection{Slit-jaw Imager}

The slit-jaw images that are reflected off the slit/prism assembly next reach 
the filter wheel. The filter wheel includes six
different filters (four for solar applications, and two for
ground testing; Table~\ref{table_sji1}). The filter wheel can be
rotated to place any one of the
filters in the beam. The NUV filters are all transmitting, whereas the
FUV filters are reflective, ensuring a different path for the NUV and
FUV SJI beams. Each of these beams includes separate reimaging and fold
mirrors. Both beams encounter the same shutter mechanism and are recorded 
on the same CCD, with one half observing the NUV SJI images, and the other
half the FUV SJI images. 

The FUV beam includes a fixed FUV bandpass filter to block light with
longer wavelengths. The NUV beam includes a Solc filter with a free
spectral range of 33\,\AA\ to reduce the
near-UV bandwidth to 3.6\,\AA\ \cite{Berger12}. The Mg {\sc ii} k and Mg {\sc ii} wing SJI filter
options are realized by combining a broader interference filter
($\approx$15\,\AA) in the filterwheel with the narrow-band Solc filter.

The four solar SJI filter options are dominated by emission from,
respectively, C {\sc ii} 1334/1335\,\AA, Si {\sc iv} 1394/1403\,\AA, Mg {\sc ii} k
2796\,\AA, and the wing of Mg {\sc ii} around 2830\,\AA. The relatively
broad passbands imply that contributions from continuum or wing
emission are significant and, depending on solar conditions, can be
dominant. Nevertheless, the bright lines are expected to contribute
significantly to the SJI images. The SJI images were chosen to provide
diagnostics over a wide temperature range, as described in
Table~\ref{table_sji1}. The C\,{\sc ii} SJI filter images may include
emission from the Fe {\sc xxi} line under flaring conditions.

To enable solid co-alignment between the various SJI and SG
channels, fiducial marks have been added to the slit. These are gaps
two pixels long along the slit, one in the top half of the CCD and one
in the bottom half of the CCD. These fiducial marks show up as dark
features in the spectra and as bright regions in the slit portion of the
SJI images.

\subsection{CCD Detector and Camera System}

The IRIS CCDs are custom designed e2v devices (CCD267) that are similar
to the AIA CCDs and are electrically compatible with the SDO camera
design.  The CCDs are back-thinned and back-illuminated with a pixel
size of 13\,$\mu$m.  Each CCD has $2061\times1056$ pixels and two
readout amplifiers that enable both halves of the CCD to be read out
simultaneously. The IRIS CCDs have been treated with e2v's proprietary 
backside--enhancement process, which eables a quantum efficiency of about
31\,\% at 1400\,\AA.  The enhancement treatment results in an annealing
pattern, visible in CCD flat--field images as an approximately
rectangular grid of lines that show sensitivity differences of order
5\,--\,10\,\% \cite{Wuelser12}.

The IRIS cameras are flight spares from SDO
(AIA and HMI). They were developed at Rutherford Appleton Laboratory and
described in more detail by \inlinecite{aia}. Each camera has four read
ports and, for SDO, each CEB was dedicated to a
four-port AIA or HMI CCD.  For IRIS, one CEB is used to read two CCD267
devices, with two ports per device.  CEB 1 reads out simultaneously the
four amplifiers from the FUV 1 and FUV 2 spectrograph CCDs.  CEB 2 reads
the two ports of the NUV SG CCD and the two ports of the slit-jaw imager
CCD.  The data is read out at two Mpixels~s$^{-1}$ through each interface, with
less than 20 electrons of read noise. The cameras communicate with the
IRIS electronics box through an IEEE\,1355/SpaceWire link. The camera has
programmable control of the waveform generator to provide different
operating modes: full frame or subfield readout, gain and pedestal
offset, flush, on-board summing, and the so-called inhibit skip
mode (which allows very rapid repeats of exposures by skipping the
full readout, but thus leaving some charge on the CCD).

\subsection{Guide Telescope and Image Stabilization System}
      \label{GTISS-instrument} 

The IRIS instrument has a guide telescope (GT) and image--stabilization
system (ISS) that are based on the systems used for TRACE and AIA. In
coordination with the spacecraft Attitude Control System (ACS), the GT allows
pointing to chosen solar targets. The ACS also can roll the
spacecraft to any angle in the range $\pm 90^{\circ}$ and provides
short--term pointing stability to arcsec or better accuracy, using the
error signals from the GT.  The ISS removes faster jitter up to about 25
Hz and provides pointing stability down to 0.05 arcsec r.m.s.

The GT consists of an achromatic refractor with an entrance filter that
has a bandpass centered at 5700\,\AA\ with a FWHM of 500\,\AA.  Four
redundant photodiodes positioned behind an occulter measure the position
of the solar limb. Intensity differences between opposite diodes are
used to derive a displacement error signal with a linear range greater
than $\pm 80$ arcsec. These analog error signals in the pitch and
yaw directions are sent to the three piezoelectric transducers 
that tilt the secondary mirror of the main telescope. The IRIS flight
software samples the GT error signals at a 32 Hz rate with a 12-bit ADC
for 0.06 arcsec resolution.  These are averaged and sent to the ACS
at 5 Hz, which controls the spacecraft pointing closed-loop to null
these errors in pitch and yaw when in Fine Sun Pointing mode.  The
ACS uses dual star trackers to control roll and also for
coarser pointing in pitch and yaw, when the GT is not in use (also known
as Inertial Sun Pointing mode).

A pair of independently rotatable optical wedges (Risley prisms) are
mounted between the entrance filter and the objective lens so that the
GT can be oriented off the optical axis of the main IRIS telescope by up
to 21 arcmin. Since the ACS continually adjusts the pointing to keep the
GT solar image centered on the four diodes, this means that the main IRIS
telescope is off-pointed from disk center to a chosen offset
location. The hollow-core wedge motors are based on HMI designs and
rotate the wedge prisms in discrete steps, so that the possible pointing
locations are a grid in polar coordinates with a maximum spacing of 16
arcsec. Moving from one position on the solar disk to another can be
done in Fine Sun Pointing mode, with the wedges moving in steps and the
ACS following the desired pointing. This typically takes about two minutes and is
less than six minutes in the longest cases. Fine-scale pointing below the
wedge grid resolution is done using the PZTs of the secondary
mirror for any pointing offset that cannot be reached because of the
discretization of the wedge motor positions.

The error signals sent to the PZTs tilt the mirror and keep the solar
image stably positioned on the slit to a precision of $<0.05$~arcsec.
Motions of the three PZTs are constrained so that the mirror only tilts
and telescope focus is maintained. The ISS is an open-loop control
system in the sense that the error signal is derived from the GT rather
than from motions of the image in the telescope focal plane. Therefore,
careful calibration is needed to match the voltage scale of the PZT
actuators with that of the GT error signals.   In addition, the
PZT actuators are part of a closed-loop servo in which the feedback
signals are derived from strain gauges incorporated into the
actuators. Using the strain gauges, accurate and repeatable image
offsets can be commanded while jitter compensation continues. The PZTs
have a linear range of $\pm80$~arcsec and a response time of 25 ms. We
reserve $\pm65$~arcsec for raster scans and the rest for a combination
of jitter correction, transients caused by momentum management or
filterwheel motion, orbital wobble, and wedge motor granularity.

The performance of the ACS and ISS has been measured on-orbit, showing
that the short--term pointing stability is usually better than 0.25
arcsec r.m.s. before correction by the ISS, and better than 0.05 arcsec
after correction. The worst transients are caused by large filterwheel
moves and momentum management of the reaction wheels. A filterwheel move
of 180$^{\circ}$ causes a rigid rotation of the spacecraft of three arcsec
over 0.5 seconds, followed by an ACS response with an overshoot of similar
magnitude lasting 10\,--\,15 seconds.  The transient is only in the direction
perpendicular to the slit (EW when the spacecraft is at $0^\circ$ roll
angle, {\it i.e.}, aligned NS) and is attenuated by the ISS to $<0.10$~arcsec.  Momentum
management transients occur when the magnetic torquers are used to
reduce reaction wheel speeds.  This occurs roughly once per orbit,
lasts one to two minutes, and causes pointing transients of a few arcsec at
most, which are slow enough that the ISS reduces the disturbance to
a level similar to that of the filterwheel transient.  Since the GT does
not sense rotation of the solar disk about its center, the GT does not
measure spacecraft roll errors.  When the science telescope is pointing
away from disk center, roll errors translate to line-of-sight pointing errors
in the azimuthal direction, with a magnitude of 17 arcsec for 1$^{\circ}$ of
roll error.  The ACS roll errors are usually less than 20 arcsec, so
these are negligible.  Occasionally, a short outage occurs during
the handover from one star tracker to the other, and this causes a
roll error that results in a pointing transient of about one arcsec
if observing at the limb.

The ACS/ISS system allows us to point anywhere on the Sun (within 21
arcminutes off disk center) with an accuracy of about five
arcsec (caused by uncertainty in the wedge motor positions).

As expected from pre-launch analyses, thermal bending of the GT
mounting causes a slowly varying pointing wobble that is in phase
with the orbital thermal conditions ({\it i.e.}, the satellite orientation
with respect to the Earth's albedo). This wobble is of order a few
arcsec over the course of one orbit and is relatively stable from
day-to-day.  The pointing wobble is nulled out by using an orbital
wobble table (see below) that includes an additional correction signal for
the PZTs that is applied as a function of time since the last ascending
node. On-orbit tests show that the correction signal
typically reduces the residual wobble to less than 0.3 arcsec.

\subsection{Mechanisms}

IRIS contains a total of eight mechanisms consisting of five distinct
types. The IRIS telescope has a front door, focus, filter wheel, 
three shutter mechanisms, and the two GT wedge motors (discussed earlier).

The front door is designed to keep the interior of the telescope
sealed from any contaminants during launch. It is based on the design
for AIA in which the door is latched shut with a high-output paraffin
actuator. The door hinge is springloaded and the door is slightly bowed to
promote the 180-degree opening when on orbit. The IRIS door
was successfully opened at the end of the bake-out period, about three weeks
after launch.

The IRIS telescope has a focus mechanism that is based on the AIA focus
mechanism, and it adjusts the position of the secondary mirror relative
to the primary along the optical axis by up to $1277 \mu$m in
2.28\,$\mu$m steps. Given the magnification factor of five, this corresponds
to approximately $57\,\mu$m change in focus per step.  The focus can be
adjusted as often as every set of exposures, but this is not necessary
for routine science observations, and is only used for some calibrations.

The filterwheel mechanism is based on the HMI focus mechanism. It
consists of a thin, brushless, DC motor manufactured by H. Magnetics to
which is added an optical encoder.  The mechanism is operated as a
stepper motor with 180 steps per revolution.  The motor has six primary
positions and a move between positions ({\it i.e.}, 30 steps per position,
Table~\ref{table_sji1}) requires approximately 0.18~seconds per position, plus
an additional 0.07~seconds of overhead.

The shutters are based on the designs for the AIA and HMI
shutters. Three shutters are installed: one each for FUV, NUV, and
SJI. The main difference between these shutters is that the FUV
shutter is significantly larger than the other two shutters so that it
can cover the two FUV CCDs. The minimum exposure times that are
supported are 112 ms for the FUV and 36 ms for both the SJI and NUV.

\subsection{Electronics}

The IRIS electronics are based on heritage from the AIA electronics
box. The IRIS electronics box (IEB) provides power and housekeeping
interfaces to the spacecraft in addition to the science-data interface
by way of two redundant IEEE 1355 SpaceWire ports. A BAe RAD6000 CPU
supports the flight software, which 
is responsible for receiving commands from the spacecraft and controls
the flow of housekeeping data and science data to the spacecraft. It
is also responsible for the interfaces to the mechanisms, the heaters,
the guide telescope, and the cameras. The portion of the flight software 
that controls the science observing program is called the instrument 
sequencer, and is described in the next section.

\section{Instrument Sequencer} 
      \label{S-sequencer} 

\subsection{Sequencer Overview}

IRIS observations are controlled by the instrument sequencer portion of
the flight software, which makes use of a series of tables, which are
prepared on the ground, checked into a database, and uploaded to the
spacecraft as needed.  Through these tables, the observational goals are
achieved by specifying the number, order, and cadence of images and
spectra to be acquired.  The tables specify how the flight software
controls the mechanisms, cameras, and the data--processing chain that
delivers the data to the spacecraft.  There are four main types of
sequencer tables (see Table~\ref{table_tables}), which have a hierarchical
structure. The lowest level are the Camera Readout Strcuture (CRS) and
Frame Definition Block (FDB) tables, which control
details for reading out the CCD.  The Frame List (FRM) controls the
timing and order in which CCD frames are acquired.  The highest level
is the Observing List or OBS table, which contains a science program
made up of various FRMs.  Observing lists are scheduled in a timeline that
is typically uploaded on a daily basis.

  \begin{table}
   \caption{IRIS sequencer tables.  The sequencer tables have a
     hierarchical structure.  OBS tables are the ``highest'' level and
     they ``call'' FRM tables, which in turn ``call'' FDB and CRS tables.}
   \label{table_tables}
    \begin{tabular}{lll}     
      \hline                   
      {\bf Table name} & {\bf Acr.} & {\bf Definition}\\
\hline

Camera Readout Structure & CRS & CCD regions (spectral lines and field
of view), \\
& & Summing modes\\
Frame Definition Block & FDB & CRS, Exposure time, Compression parameters,\\
& & Lookup Table\\
Frame List&FRM&Time, FDB, Repeat Count, PZT offset (raster\\
& & scan), Focus, Slitjaw bandpass, Cadence\\
Observing List & OBS&Time, FRM, Repeat Count, Cadence, PZT offset\\
\hline
    \end{tabular}
 \end{table}

The CRS, FDB, FRM, and OBS tables contain all the information required to 
acquire an image or spectrum:

\begin{enumerate}
\item CRS table data controls which data are read out from the CCD and
  sent on to the data compression/high rate interface (DC/HRI) board.

\item The FDB configures the shutter to obtain the desired exposure time. 

\item At the time specified by the FRM and OBS the following
  sequence is started:
\begin{itemize}
\item the mechanisms are moved (if required/desired): focus position, PZT offsets, filterwheel.
\item the shutter is opened.
\item after exposure time, the shutter is closed.
\item data are then read from the CCDs by the cameras and sent to the
  DC/HRI board,
\item the compression parameters in the FDB are used by the DC/HRI to compress the images before sending them to spacecraft memory.
\end{itemize}
\end{enumerate}

\begin{figure}    
  \centerline{\includegraphics[width=1\textwidth,clip=]{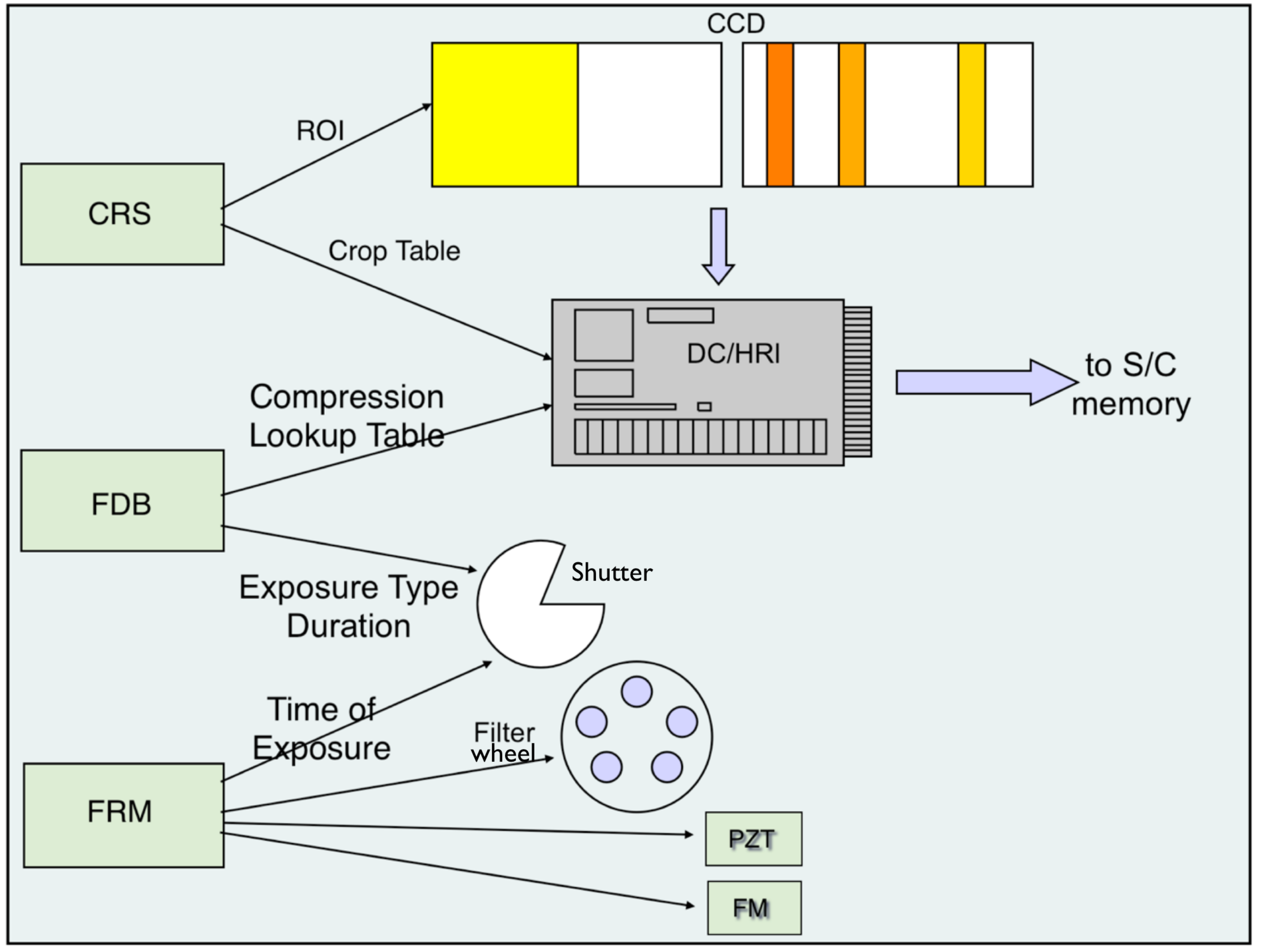}}
              \caption{How IRIS images and spectra are
              taken and controlled by the Camera Readout Structure
              (CRS), Frame Definition Block (FDB) and FrameList (FRM).}
   \label{fig_sequence}
    \end{figure}

The flight software ensures that the shutters are not opened until all
mechanisms have stopped moving and settled. In addition, the exposures
controlled by all three shutters are synchronized so that they all end at
the same time, followed by the camera readout. Depending on a
parameter set in the OBS tables, the camera readout can
be done in parallel for both cameras, or sequentially (thus reducing
low-level noise cause by electronic interference from simultaneous readouts). The typical sequence of events is illustrated in Figure~\ref{fig_sequence}.

The CEBs allow for on-board summing, {\it i.e.}, before digitization in the
A/D converter. This is controlled by the CRS tables, which allow any
combination of 1$\times$, 2$\times$, 4$\times$, 8$\times$ spectral summing and 1$\times$, 2$\times$, 4$\times$ spatial
summing. The sequencer software also allows the readout of small
regions of interest on any of the CCDs. This speeds up readout of the
CCDs and reduces the data volume that is downlinked. The latter is
also accomplished by on-board compression. IRIS follows the same
compression approach as AIA. It has one data
compression/high-rate interface (DC/HRI) card that performs data compression
in hardware and then transmits the compressed data to the spacecraft
interface and the solid--state memory board (SSMB, which has a capacity
of 48 Gbits). The primary algorithm is lossless Rice compression, which
typically achieves a factor of two reduction in data volume for UV
spectra and images. A second, lossy algorithm is based on a look-up
table, which is configurable from the ground. The baseline tables are
variations of  scaled square-root function, similar to those used 
for HMI and AIA. This reduces the data volume by another factor of
about two, depending on the type of images/spectra to which it is applied.

\subsection{Timing}

The IRIS sequencer software has been designed to take spectra and images
at a very high cadence to accommodate the science goals of IRIS which
call for two-second spectral and five-second imaging cadence.  The ``Take Picture''
command in the flight software accomplishes the key operations that
enable the appropriate IRIS data to be acquired and is illustrated in
Figure~\ref{fig_timing}.  Each activity requires a certain minimum amount of
time, the timing of which defines the the minimum (or fastest) cadence that 
IRIS can obtain. There are three groups of events that are optimized for
fastest cadence operations and they are in part sequential and in
part parallel (Figure~\ref{fig_timing}):

  \begin{figure}    
  \centerline{\includegraphics[width=1\textwidth,clip=]{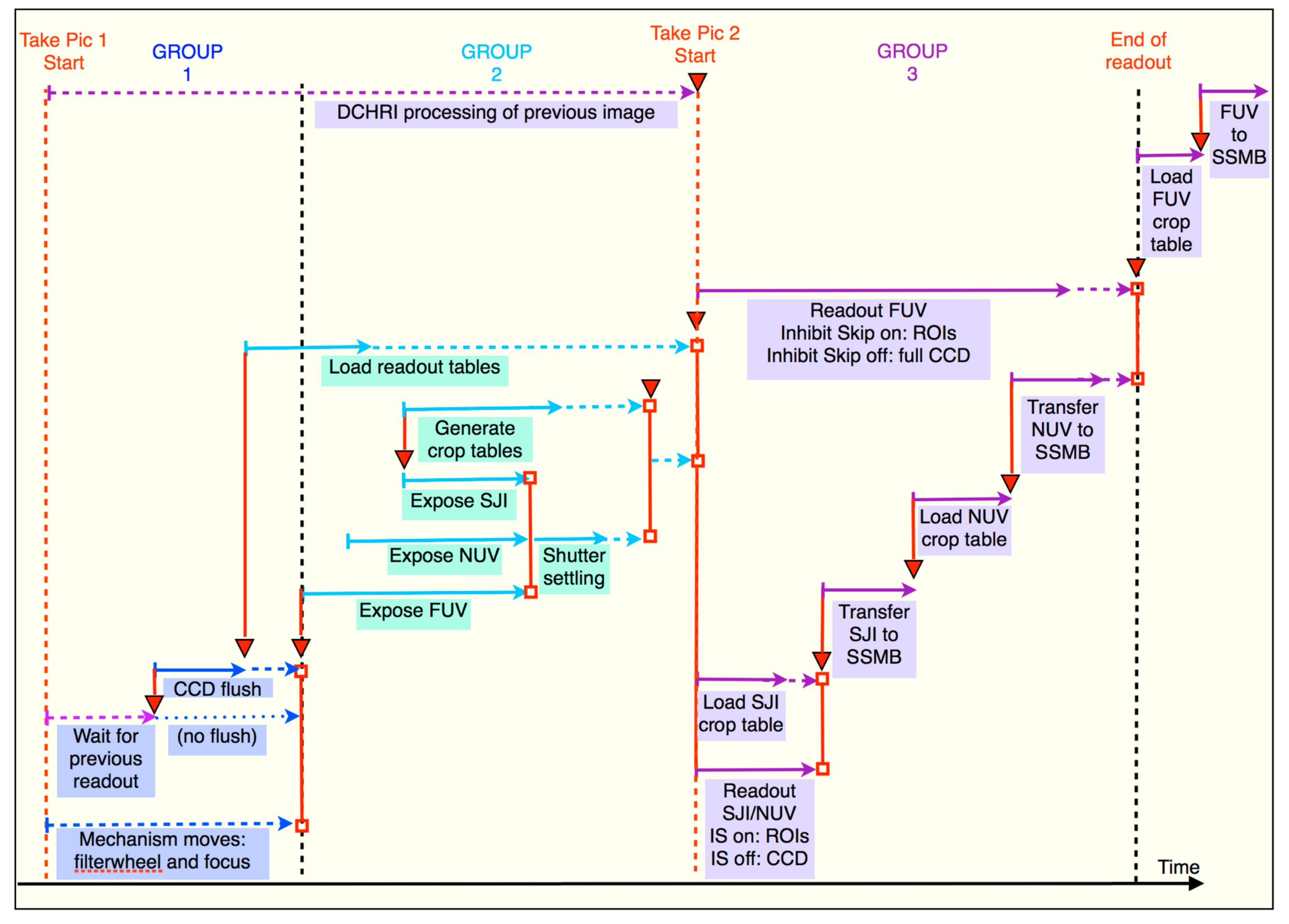}}
              \caption{The sequence of events involved
                in taking and processing a ``picture''.}
   \label{fig_timing}
   \end{figure}

\begin{enumerate}
\item mechanism moves, waiting for readout of previous image(s) and CCD flush,
\item starts with the longest exposure time and ends with the start of
  the readout and/or DC/HRI processing of the previous image,
\item camera readout and processing of the images in the DC/HRI. As
  soon as the readout starts, the mechanism moves and the next ``Take
  Picture'' sequence can start.
\end{enumerate}

To optimize the timing of observing sequences, a ground-based ``Table
Creator'' tool has been developed that simulates the behavior of
sequencer software and hardware and adjusts the timing in the OBS and
FRM tables so the instrument runs as fast as possible.

\subsection{Default Tables}

To minimize complexity of table management and testing, as well as
operations, a large number of sequences have
been calculated and are available to science planners. The numbering
scheme of the default OBS tables reflects their scientific goal as
illustrated in Tables~\ref{table_default1}, \ref{table_default2}, and \ref{table_default3}. 
There are about 50 basic observing modes (Table~\ref{table_default1}), which
vary from dense, coarse, and sparse rasters to sit-and-stare sequences
of varying sizes and field-of-view choices. These basic sequences are
also available with variations in SJI cadence and wavelength selection, exposure
times, wavelength coverage, {\it etc}. These are listed
in Tables~\ref{table_default2} and \ref{table_default3}. 
The numbering scheme is such that the same basic
observing-mode properties are maintained in the last two digits of the
OBS-ID, whereas leading digits indicate various choices for exposure
times, SJI cadence, etc. The numbering scheme thus acts similarly to a
binary mask with the digits listed in Tables~\ref{table_default2} and \ref{table_default3} acting as
“bits” switching options on and off.  For example, OBS table 20113644 defines 
a large ($126\times120$ arcsec) raster with coarse (two arcsec) steps, a medium 
linelist with lossy compression, 30~second SJI images in Si {\sc iv} and
with two by two 
spatial binning and $2\times$ spectral binning.  
Note that the leading two digits of any OBS-ID are the ``version
number'' (not shown in the example), {\it e.g}, 3820113644 is part of table
version 3.8. OBS-IDs starting with ``42'' are calibration sequences. 

The basic raster modes include:
\begin{itemize}
\item dense rasters: step size of the raster is equal to the slit width (0.33 arcsec). 

\item sparse or coarse rasters: step size of each raster location is larger than
  the slit width. Allows rapid scans of much larger areas, {\it e.g.}, for
  flare or CME watch programs.

\item sit-and-stare (fixed slit mode): no rastering.

\item multi-point dense/sparse rasters: involves rasters with a small
  number of dwelling locations to study propagation of waves, etc.
\end{itemize}

The smallest possible stepsize is 0.054 arcsec, which is also the
granularity with which solar rotation is tracked with IRIS.

\section{Operations} 
      \label{S-ops} 

\subsection{Overview}
The IRIS polar, Sun-synchronous orbit is similar to that of TRACE and
{\it Hinode} and allows for 7\,--\,8 months of continuous observations per year,
with strong atmospheric absorption occurring during the
November\,--\,February time frame when the Sun is seen by IRIS at heights
below $\approx200$ km (FUV) and $\approx50$ km (NUV) above the Earth's
surface. IRIS data are downlinked to X-band antennas in Svalbard, Norway
($\approx$8\,--\,9 passes\,day$^{-1}$), Alaska ($\approx$3\,--\,4 passes\,day$^{-1}$) and Wallops
($\approx$1\,--\,2 passes\,day$^{-1}$). As a result, IRIS has an average data rate of
0.7 Mbit~s$^{-1}$, {\it i.e.}, about 60 Gbits~day$^{-1}$ (compressed). Data (nominally 14
bits~pixel$^{-1}$) are compressed onboard using Rice and lossy compression,
nominally to about 3\,--\,4 bits~pixel$^{-1}$.  Onboard memory is 48 Gbits,
allowing storage of more data-intensive observing sequences that can be
downlinked later over the course of several orbits.

The high data rate, short exposure times, and flexible rastering
schemes allow rapid scans of small regions on the Sun at very high
spatial resolution of order 0.33\,--\,0.4 arcsec. The baseline cadence 
is five seconds for slit-jaw images, and three seconds for 6 spectral windows of strong, bright lines.

IRIS operations are similar to TRACE and
Hinode, with observing programs uploaded five times per week and 
data made publicly available within a few days of the
observations. Coordination with {\it Hinode}, SDO and a variety of
ground-based observatories is a high priority.

\subsection{Timeline}

IRIS operations are controlled by a timeline, which contains a list of spacecraft and instrument commands. The timeline is used by a science planner to perform the following operations:
\begin{itemize}
\item point the telescope to a position on (or off) the solar disk
  within 21 arcminutes of disk center.

\item switch solar tracking on or off. When solar rotation tracking is
  on, the pointing of the telescope is continuously adjusted to
  compensate for the rotation of the Sun so that the same region on
  the Sun is kept within the field of view. 

\item correct the pointing of the telescope to compensate for orbital wobble introduced by thermal flexing between the guide telescope and the main IRIS telescope. This is done by using an orbital wobble table (Section~\ref{GTISS-instrument}).

\item roll the IRIS telescope from its nominal direction (slit
  parallel with the N--S direction on the Sun) by an angle between
  $-90$ and $+90$ degrees from solar north.

\item perform an upload of OBS and dependent tables.

\item execute (start and stop) a number of previously uploaded OBS
  tables. 

\item set the global automatic exposure control (AEC) parameters.

\item switch on and off transmission of data through an X-band downlink.

\item close or open the ISS loop (to remove jitter).
\end{itemize}

The timeline is generated using the timeline tool, which is based on
the TRACE timeline tool. Timelines are uploaded to IRIS every weekday.
The Monday through Thursday timelines cover a planning period of one
full day from 0400 UT to 0400 UT. The Friday timeline covers a time period from 0400 UT Saturday to 0400 UT Tuesday.

%

\subsection{Roll Restrictions}
The nominal direction of the IRIS slit is parallel with the solar
North--South direction (roll angle\,=\,0$^\circ$). However, IRIS is capable
of operating with the spacecraft rolled at any angle between $\pm90^\circ$ 
from solar North. This enables observing programs in which
the slit is parallel with the solar limb at any position along the limb
(from the Equator to the poles).

Operations under rolled conditions have two operational impacts or restrictions:
\begin{itemize}
\item reduced downlink data rate, caused by the directional X-band antenna
  no longer pointing “straight down to Earth” for non-zero roll
  angles. Some roll angles will lead to significant reduction of
  positive link margin and result in shorter downlink passes.

\item certain roll angles are forbidden twice per month (first and last
  quarter of the Moon). The IRIS attitude is controlled by two star trackers
  at opposite ends of the spacecraft. During two periods around the
  first and last quarter of the Moon there are certain roll angles for
  which the Earth is in one star tracker and the Moon is in the other star
  tracker. When that occurs the star tracker CCDs may become saturated
  and attitude control is no longer possible, which causes IRIS to
  automatically transition into a safe ``non-science gathering mode.'' To avoid this, certain
  roll angles are excluded twice a month.
\end{itemize}

\subsection{AEC Operations}

One of the science goals of IRIS is to study how flares and CMEs 
are triggered ({\it e.g.}, by flux emergence). Obtaining spectra and images
 during the impulsive phase of flares is an important aspect of these
 types of studies. The occurrence of flares presents a major challenge
for a slit-based spectrograph: intensities of spectra vary strongly spatially and
temporally, which can lead to overexposure and saturation of some of
the spectra. Automatic exposure control (AEC) based on spectra taken as part
of a raster scan across the solar disk is not an efficient way of
avoiding overexposure because the AEC input data lags behind the
rapidly evolving brightness in a flare environment in time and space.
The AEC approach for IRIS circumvents this problem by having the AEC
for all detectors (spectra and images) be based on intensities measured
in the SJI images. By design, IRIS does not switch to a different, specific 
observing  program after flares are detected, so the current observing program
will continue to run, but with
SJI-based AEC for spectra and images.

The general IRIS AEC philosophy is not focused on trying to capture very weak signals. Instead it is similar to that of SDO/AIA, which is based on making sure over-exposed data do not occur too often. In other words, the AEC is focused on preventing saturation during flares.

  \begin{table}
   \caption{IRIS Automatic Exposure Control global parameters}
   \label{table_aec}
    \begin{tabular}{lll}     
      \hline                   
      {\bf Name} & {\bf Acr.} & {\bf Definition}\\
\hline
Flare Flag & FF & Controls whether IRIS goes into flare AEC mode or\\
& & not. If FF\,=\,1, SJI images control exposure times of spectra\\
AEC Flag &AF & Controls whether the AEC software  \\
& & is used to change exposure times (AF\,=\,1) or not\\
AEC Lapse & ALT & Exposure times are set to default in the FDB\\
Time & & if last SJI was taken more than ALT seconds ago\\
Event Flag & EF & Keeps track of whether flaring event has occurred or
not\\
Event Lapse & ELT & After ELT seconds, the event flag is set back to
zero\\
Time & & \\
Event Enable & EEM& Sets which of 16 AEC tables is used to determine whether\\
Mask& & an event has occurred\\
    \end{tabular}
 \end{table}

Most of the AEC operations are set at the FRM and FDB level. By
default the IRIS timeline will set the AEC flag to zero, {\it i.e.}, disable AEC
operations (see Table~\ref{table_aec}). If the planner sets AEC to one for an observing program, the
onboard software will automatically determine for each SJI image set
in the Event Enable Mask (EEM) whether a flaring event (Event Flag\,=\,1)
has occurred or not. The EF will be set to one by the onboard software
when the computed AEC exposure times drop below a specified threshold
exposure time. This threshold exposure time is set in the AEC
tables. When EF is set to one, AEC of spectra and images exposure times are controlled
by the AEC options listed in the FRM list. After EF was set to one, the instrument software sets EF\,=\,0 when
a flare condition has not been satisfied for the time specified by
ELT. Similarly, AEC operations are suspended if no SJI images were
taken within ALT seconds.

\section{Calibration}
\label{S-calibration}

\subsection{CCD Characterization}
\label{S-CCD}

The four CCDs are read out by two different CEBs. The gain for the FUV CEB has been set high to six
electrons\,DN$^{-1}$. All SJI channels and the NUV use a lower gain setting of
18 electrons\,DN$^{-1}$. All NUV photons create one electron/hole pair on the
detector, {\it i.e.}, one electron\,photon$^{-1}$. This means that for NUV spectra and
NUV SJI channels 2796 and 2832 there are about 18 photons\,DN$^{-1}$. For FUV
photons about 1.5 electrons\,photon$^{-1}$ are created on average. This means
that for FUV spectra there are about four photons\,DN$^{-1}$, while for FUV SJI
channels 1330 and 1400 there are about 12 photons\,DN$^{-1}$.

The dark frame levels for IRIS have two components: a pedestal level
[$P$] set electronically, and the dark current.  Each CCD read port
exhibits slightly different behavior.  Extensive tests pre- and
post-launch have led to the following model for the total dark
level [$D$] in read port $j$:

\begin{equation}
D_j =  P_j[T_{{\rm CEB}j}(t-\delta t_j)] +
       {\rm e}^{(a_j + b_j T_{{\rm CCD}j})} n_x n_y  t_{\rm int}
       + \Delta D_j(x, n_x, n_y, t_{\rm int})
\end{equation}

Here, $P_j$ is the pedestal level in read port $j$, a function of CEB temperature [$T_{{\rm CEB}j}$] for that port,
time lagged by  $\delta_t$. The second term models the average dark-current
rate, which is the product of an exponential dependence on the CCD temperature
[$T_{{\rm CCD}j}$], the amount of on-chip summing [$n_x n_y$] and the
time between CCD reads ({\it i.e.}, the dark-current integration time) [$t_{\rm int}$].
A final term [$\Delta D$] models the change in shape of the dark in the
wavelength ($x$) direction as $t_{\rm int}$ and summing are increased,
from flat for $t_{\rm int} \approx 0$\,seconds and $1\,\times\,1$ summing, to
roughly bilinear in $x$, sloping gradually and then more rapidly away from
the read-out point, with an end-to-end amplitude of $\Delta D \approx 10$
DN for $n_x n_y = 32$ and FUV ports.

In practice, the $D_j$ function is computed for each port, and added
to the appropriate portion of an averaged, cleaned ``basal" dark.  This
``basal" dark was constructed from 19 $t_{\rm int} \approx 0$\,seconds
images, which had particle hits and (some) hot pixels removed;
the average value in each read-port zone was then set to zero.  Analysis
of 152 full-frame $1\times1$ darks indicates average errors at
$t_{\rm int} = 30$\,seconds of $<0.12\,\pm\,0.12$ DN in the zero point (worst
case) for the FUV ports, and $<0.08\,\pm\,0.03$ DN for the NUV/SJI
ports.  Scatter in the residual RMS(dark - model dark) is $< 3.1$
DN (FUV ports) and $< 1.2$ DN (NUV/SJI), and essentially
reflects the camera readout noise.  Results deteriorate
somewhat as summing increases:  for $n_x n_y = 8$,
average offsets are $<1.6\,\pm\,2.5$ DN for FUV, and  $<0.16\,\pm\,0.82$ DN
(NUV/SJI), with RMS residuals of $<$ 3.3 DN (FUV) and $<$ 1.2 DN (NUV/SJI).
These calibrations will be revisited and refined periodically as
more data are taken and more operational temperatures are sampled
over the course of the spacecraft's thermal ``year".


On orbit, it was discovered that when both CEBs are read out
simultaneously, a read-interference noise pattern is superimposed
on the resulting data.  While it does not always occupy the entire
read-port zone spatially, the pattern, where it is present, is
approximately constant in the spatial direction for a given readout
and read port.  The portions of the CCD showing the pattern show
mirror symmetry across the two read-port zones of a given chip, and
both CCDs show identical coverage.  The pattern's wavelength structure
is complex and varies in shape and period from readout to readout,
with a typical period of about 340 pixels ($1\times1$ summing).   The
period is constant for all ports of a given readout, though its
shape varies from port to port.  One port typically has the highest
amplitude pattern (in FUV SG $\approx \pm$ 3 DN), with other ports showing
decreasing amplitudes by successive factors of $\approx$ two, and simpler
wavelength structure.   This pattern is barely visible in
the high contrast, high S/N NUV spectra and all SJI images. It is more
apparent in FUV spectra which have higher gain and typically have
lower counts. To mitigate this pattern noise, IRIS planners can choose
to use non-simultaneous reads which eliminates the pattern completely
at the cost of lower cadence. The IRIS team is also working on
software to model and remove the readout pattern.

\subsection{Flat Field}

Flat fielding of images acquired with IRIS CCDs is required to remove the effects of:
\begin{itemize}
\item vignetting,
\item the semi-regular laser anneal pattern visible at the 5\,--\,10\,\% level in CCD images,
\item potential dose accumulation burn-in,
\item dust and artifacts on the slit and detector,
\item the intensity pattern due to irregularities in the slit width.
\end{itemize}

Flat fields for IRIS are constructed from solar observations for the NUV and FUV spectrographs and the slit-jaw imagers with the same intent, to extract structure due to the instrument from structure due to the Sun, although two different approaches are used.

\subsubsection{Strategy for the Spectrograph Flats}
The spectrograph flat fields are produced by extracting the spectral profile from spectrograph images which have been sufficiently averaged.  The emission lines of the FUV and the Mg {\sc ii} line cores show high-contrast features and non-uniform line profiles, even in the quiet Sun.  A spectrum which is spatially smooth and spectrally uniform is obtained by taking a coarse raster of the disk-center quiet Sun, during which the telescope is defocused and slewed during the exposures to provide additional smoothing, then the images in the sequence are averaged together.  At least 200 images are required to build a good intermediate flat field.

To extract the spectral profile the intermediate flat field is up-sampled by padding the Fourier transform of the image.  This image is warped using the geometry correction determined for each spectral window to make the spectral and fiducial lines straight with respect to the grid of pixels.  Averages are taken along the spatial and spectral dimension.  These averages are reversed through the geometry correction, resampled back down to the original image size, and divided from the original spectrum to produce the final flat field.  The spatial flat field (average along the spectral dimension) is reserved so that it can be shifted and applied to properly compensate for slight thermal shifts of the spectrum on the detector.

Exposure times of 15 seconds are used for each individual image for both the NUV and FUV flat-field observations.  For the FUV spectra, this ensures that the bright lines have enough signal, however the faint lines and continuum do not rise above the level of the read noise.

Analysis of the FUV spectra has revealed a faint, smooth stray-light background that needs to be removed before flat-fielding. 
This background does not contain the CCD annealing pattern that is present at blue and UV
wavelengths, indicating that it is a visible-light leak,
whose photons penetrate more deeply into the CCD.  A gain of $5.6\pm0.2$
photons\,DN$^{-1}$ has been measured for the stray-light background in long
exposures taken off the solar limb, which is comparable with the gain
measured with the blue LEDs and consistent with the statistics
expected for visible photons.  Analysis of the off-limb
data from the other channels of IRIS does not indicate a similar stray-light
component.  The FUV background is of order 0.5 DN\,s$^{-1}$ at most and has been found to drop
with distance from Sun center to about 0.15 DN\,s$^{-1}$. This
drop-off varies with position angle of the slit relative to the Sun,
which suggests that the background does not enter through the slit.  
It is more likely due to a low-level light leak in the labyrinth
structure of the slit prism mount.  The intensity of the
background also depends on the position of the SJI filter wheel.
It is slightly higher with an NUV filter in place, probably because
these filters reflect some visible light back to the slit prism area. The background has been characterized 
and is being removed before the FUV spectrograph flat fields are applied in our Level 1.5
calibration.

\subsubsection{Strategy for the Slit-Jaw Imagers}
Flat fields for the slit-jaw imagers are constructed using the
technique presented by \inlinecite{Chae04}, which makes use of
relatively shifted solar images to extract the stationary
flat-field pattern of the instrument from the moving solar scene.  For
this implementation we have selected a Reuleaux triangle dither
pattern of 20 arcsec containing 15 pointings.  This way the dither can
be accomplished on the short timescale of the solar chromosphere.  The
results of multiple pointings are combined to produce the final flat
field.  Regions of bright, but quiet, plage provide sufficient counts
for the FUV.  For the filters with high contrast (FUV and NUV core),
images are taken with the telescope defocused so that a more uniform coverage of the CCD is possible, while images for the lower-contrast NUV wing and glass filters are taken in-focus to ensure the granulation pattern of the quiet Sun can be accurately extracted.

The Chae method significantly decreases the amount of time spent
obtaining flat-field images.  Each wavelength of the SJI uses
different optics, causing a change in the vignette pattern. In
addition the CCD has a slightly different gain response at each
wavelength, necessitating the use of separate flat fields for each filter.  Using the Chae method, only $15\times3$ flats are necessary for each filter, whereas an averaging method might take hundreds of images to sufficiently smooth over the solar structure.

The fiducial marks in the spectra and the slit in the SJI images are
important for calibration and are therefore seamlessly removed from
the flat-field by replacing the pixels around each with the pre-launch flat-field images obtained using a mercury lamp filtered to 2537\,\AA.


Flat-field calibration data will be acquired regularly to track
changes ({\it e.g.}, during eclipse season, or because of burn-in) with
appropriate changes made to the calibrated Level 2 data.

\subsection{Optical Performance}

\subsubsection{Focus}

Focus is adjustable by moving the telescope secondary mirror in
2.28\,$\mu$m increments, over a range from $-210$ to +350 steps (a full
range of 1277\,$\mu$m).  The
best position of telescope focus was determined by analyzing multiple
sequences of solar images at different focus positions. The results are
summarized in Table\,\ref{tab_focus}. Uncertainties are less than one
step, except for the FUV spectrograph. The profusion of
transition-region explosive events makes the quantitative estimate of focus
difficult for the FUV, but visually it is near $-115$.  Depth of focus is
approximately $\pm 5$ steps. IRIS runs routinely at the designated best
instrument focus of $-115$. During eclipse season, the focus changes
slightly to values around $-121$.

\begin{table}
  \caption{Best focus position (in steps) for each IRIS science channel} 
  \label{tab_focus}
  \begin{tabular}{ll}
     \hline
      Channel & Focus position\\ 
\hline
      SJI 1330\,\AA & -114.4 \\
      SJI 1400\,\AA & -114.5 \\
      SJI 2796\,\AA & -113.6 \\
      SJI 2832\,\AA & -111.8 \\
      NUV SG        & -116.3 \\
      FUV SG        & $\approx -115$\\
\hline
   \end{tabular}
\end{table}


\subsubsection{Point Spread Function and Spatial Resolution}

\begin{figure}    
  \centerline{\includegraphics[width=0.5\textwidth,clip=]{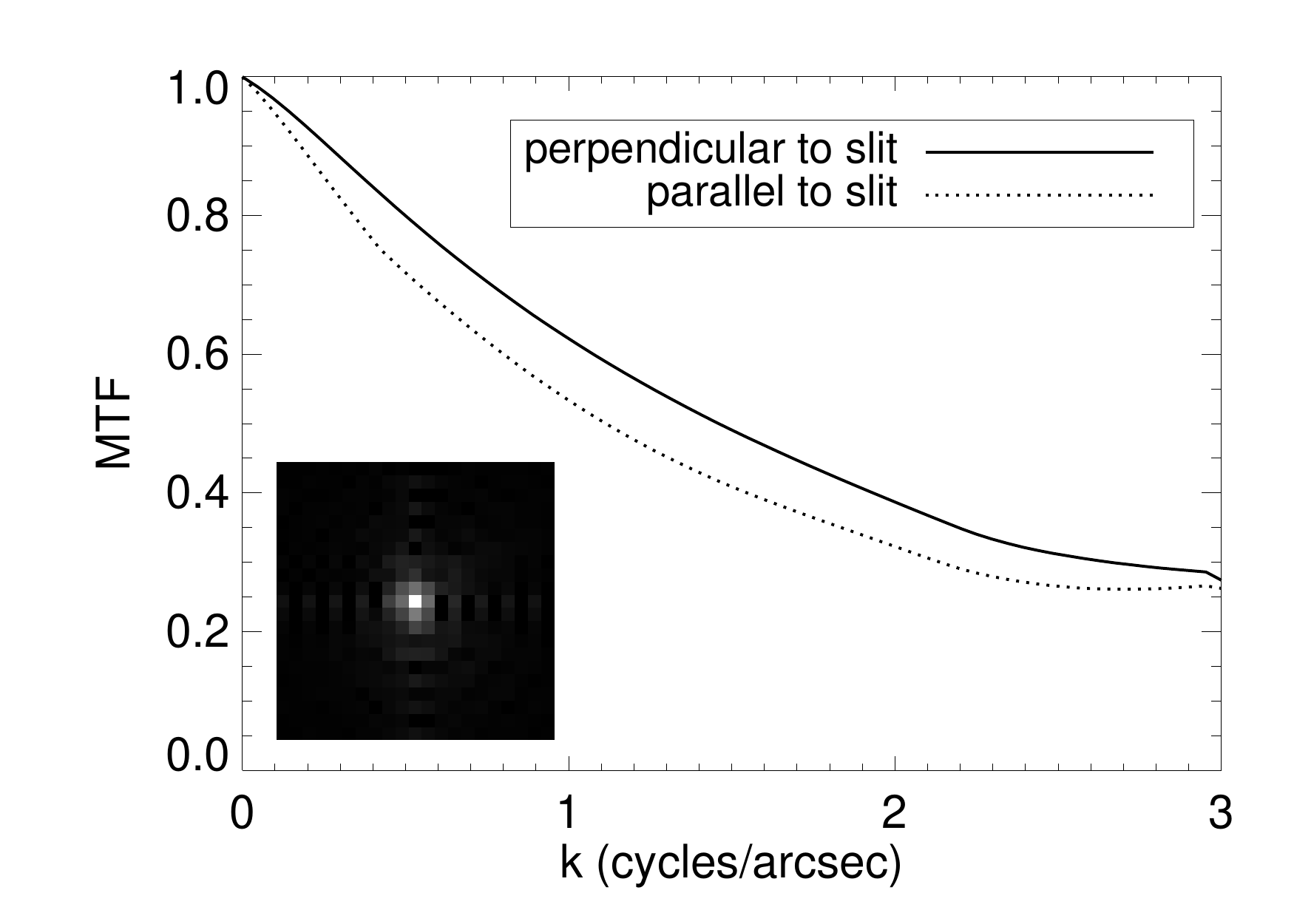}\includegraphics[width=0.5\textwidth,clip=]{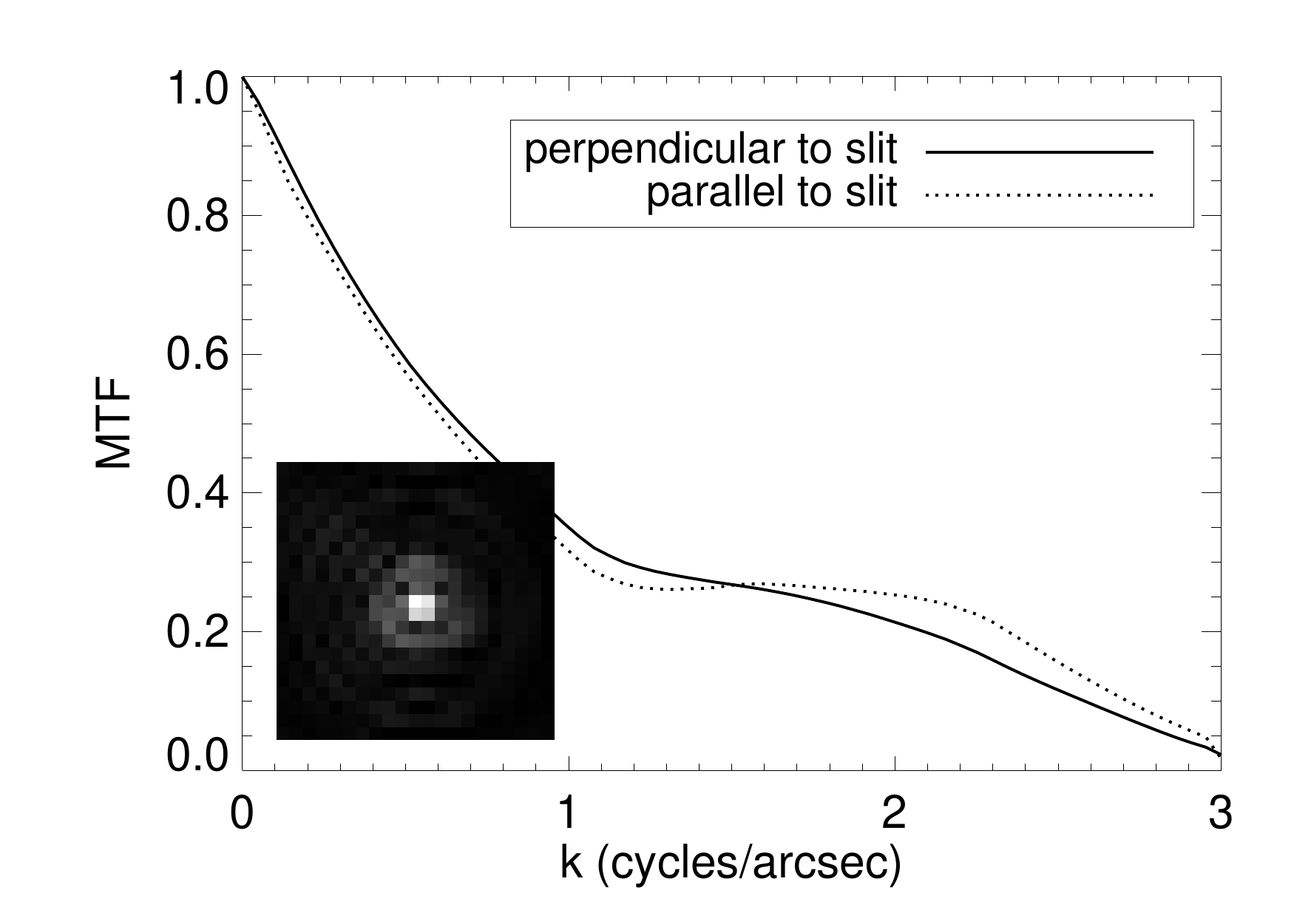}}
\caption{MTF of the IRIS FUV (left) and NUV (right) SJI, derived from phase-diversity analysis of a series of exposures adjusting through focus at $1330$\,\AA. The PSF, inset, has been square-root scaled to bring out faint features.}
\label{fig:mtf1330}
  \end{figure}

The spatial resolution of IRIS is very good in both the spectrograph
and SJI. Figure \ref{fig:mtf1330} shows point-spread functions and modulation
transfer functions for the FUV 1330 and NUV 2832
channels of the IRIS SJI. These were derived from on-orbit focus tests
using a phase-diversity code described by \inlinecite{Lofdahl1994}. The code
was run on a small region centered in the lower left portion of each
detector (coordinates 150, 250). The NUV SJI is operating near the
diffraction limit, with an MTF dip at $\approx 1$ cycles\,arcsec$^{-1}$ due
to the telescope central obscuration. The MTF falls to 10\% at 2.6
cycles\,arcsec$^{-1}$, which exceeds the resolution requirement for the
instrument.  The NUV images have a 2-pixel wide
 core containing $\approx\,42$\,\% of the light.

The figures show that FUV images are nearly pixel-limited, with the
MTF above 0.25 all the way to the Nyquist frequency: the PSF core is a
single bright pixel containing 34\,\% of the light and the
 $3\times3$ brightest pixels have 70\,\%.

\subsubsection{Spectral Resolution}

\begin{figure}    
  \centerline{\includegraphics[width=0.5\textwidth,clip=]{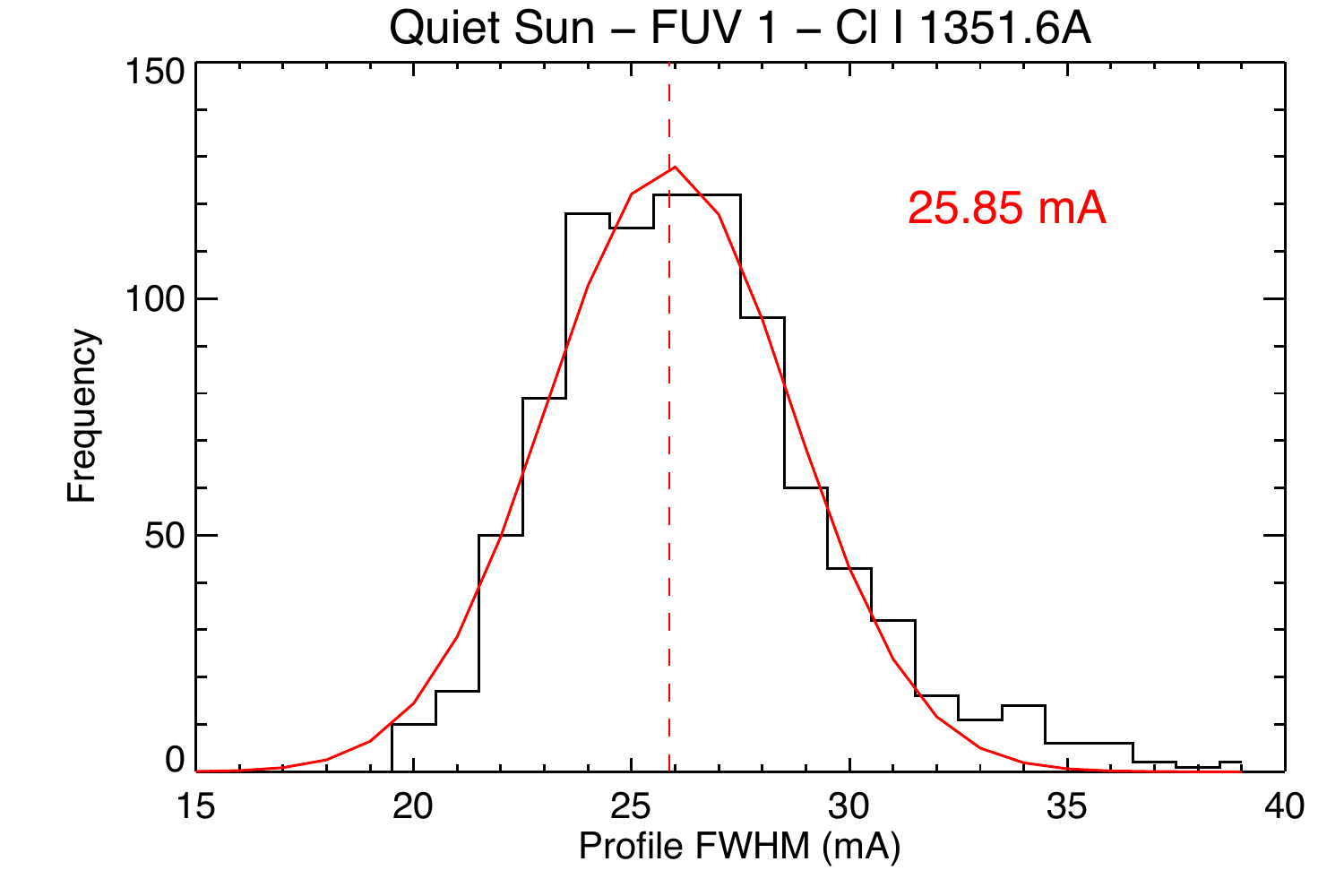}\includegraphics[width=0.5\textwidth,clip=]{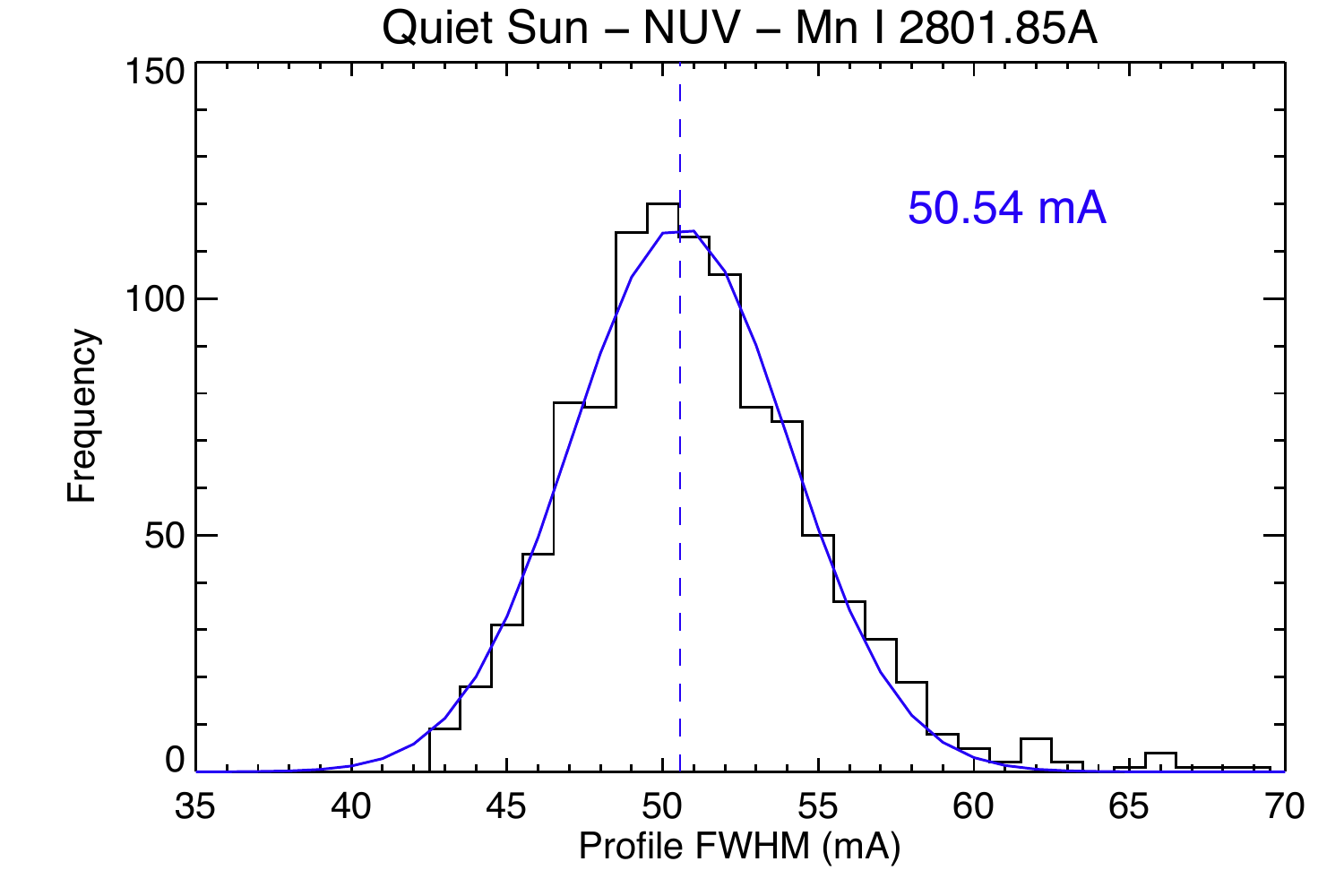}}
              \caption{Histogram of the FWHM of the O {\sc i} 1351.6\,\AA\ and
              Mn {\sc i} 2801.85\,\AA\ lines in a quiet-Sun region indicates
              that the effective spectral resolution is limited by the
            Nyquist criterion. The dashed lines show the averages of
            the distributions.}
   \label{fig_specres}
   \end{figure}

Preliminary measurements of the spectral resolution of IRIS show it to
be excellent, and essentially limited by the Nyquist criterion. We
analyzed quiet-Sun spectra taken on 20\,Aug\,2013 and used single
Gaussian fits to the O {\sc i} 1355.6\,\AA\ and Mn {\sc i} 2801.85\,\AA\ lines to
determine the FWHM of these lines for 1000 positions along the
slit. We find that many profiles of these lines are very narrow with
average FWHM of 25.85\,m\AA\ in the FUV and 50.54\,m\AA\ in the
NUV, {\it i.e.}, below and around the Nyquist criterion. This is
illustrated in Figure~\ref{fig_specres}, which shows histograms of the
width measured in these neutral lines.

\subsubsection{Plate Scale}

The spatial plate scales for the SJI shown in Table~\ref{table_sji1}
were determined using a meridional scan, and then
  assembling all the images into a North--South mosaic across the full disk.  In that
  mosaic, the solar diameter in pixels is compared with the known solar
  diameter, corrected for the distance to the Sun at the time of the
  observations. The SG plate scales (Table~\ref{table_sg}) were determined 
by matching the distance between the two fiducial marks on the SJI and
SG data and applying the resampling factor to the previously
determined SJI plate scales. Note that at data processing Level 1.5
(and higher) all data are placed onto a common platescale of 0.16635 arcsec\,pixel$^{-1}$.

The spectral plate scales (Table~\ref{table_sg}) are average values
across the detector.  These prelaunch values agree very well with the
averages from post-launch calibration. Note that the actual spectral
plate scale is wavelength dependent.  This non-linearity in the plate
scale is corrected in Level 1.5 data (and higher) where the data are
interpolated onto the average plate-scale grid.


\subsection{Compression}
Compression of IRIS data for efficient transmission to the ground is
accomplished through two methods:  lossless Rice compression
\cite{Rice71} and optional lossy compression using lookup tables (LUTs).

Rice compression takes advantage of the relative smoothness of data,
by taking a running difference and encoding values in two parts.  The
rapidly changing, least significant bits are represented directly and
the slowly changing most significant bits are compressed by the
running difference.  Data values in an image can be losslessly
compressed by about a factor of two using this method.  The Rice K value selects the number of least-significant-bits to be directly represented.  For IRIS, the K-value for a given observation is set in the FDB during the generation of observing tables.  The K-value for most efficient compression depends on the structure in the image and the exposure time.  The best K-values have been determined for a variety of images and spectra.

Additional compression of the IRIS data can be achieved by applying a
lossy algorithm using a lookup table (LUT).  For IRIS we make use of
the LUTs generated for AIA and HMI data from the SDO mission.  These
compression tables consist of a linear ramp at low values and
transitions to a square-root function at approximately the level of the
read noise.  This way low counts are not compressed, but higher counts
are compressed more and more toward the top of the table.  Compression
with a square-root function introduces lookup error in a constant
proportion to the photon-counting noise, resulting in a constant signal-to-noise ratio for each table.  The calculated effective ratios of the
LUT error to photon+read noise, based on the
noise level expected from a given detector, are tabulated for all the
LUTs available for IRIS in Table \ref{tbl:lut_cr}.  The values shown are
calculated for the CCD pedestal levels that were adjusted after launch.

\begin{table}[t!]
\caption{Effective S/N Ratios for AIA and HMI LUTs}
\label{tbl:lut_cr}
\begin{tabular}{clcc}
\hline
ID & Name & FUV ratio (low gain) & NUV ratio \\
\hline
0 & No Compression & 0.0 & 0.0 \\
1 & Inverse & 0.0 & 0.0 \\
2 & hmi c3 plin & 0.28 & 0.32 \\
3 & aia csnr 025 & 0.91 & 1.0 \\
4 & aia csnr 050 & 1.8 & 2.1 \\
5 & aia csnr 100 & 3.7 & 4.2 \\
6 & hmi c3 5 plin & 0.40 & 0.46 \\
7 & hmi c4 plin & 0.57 & 0.65 \\
8 & hmi c4 5 plin & 0.81 & 0.92 \\
9 & aia csnr 200 & 7.6 & 8.6 \\
\hline
\end{tabular}
\end{table}

\subsection{Pointing Stability}

\begin{figure}[h]    
  \centerline{\includegraphics[width=0.5\textwidth,clip=]{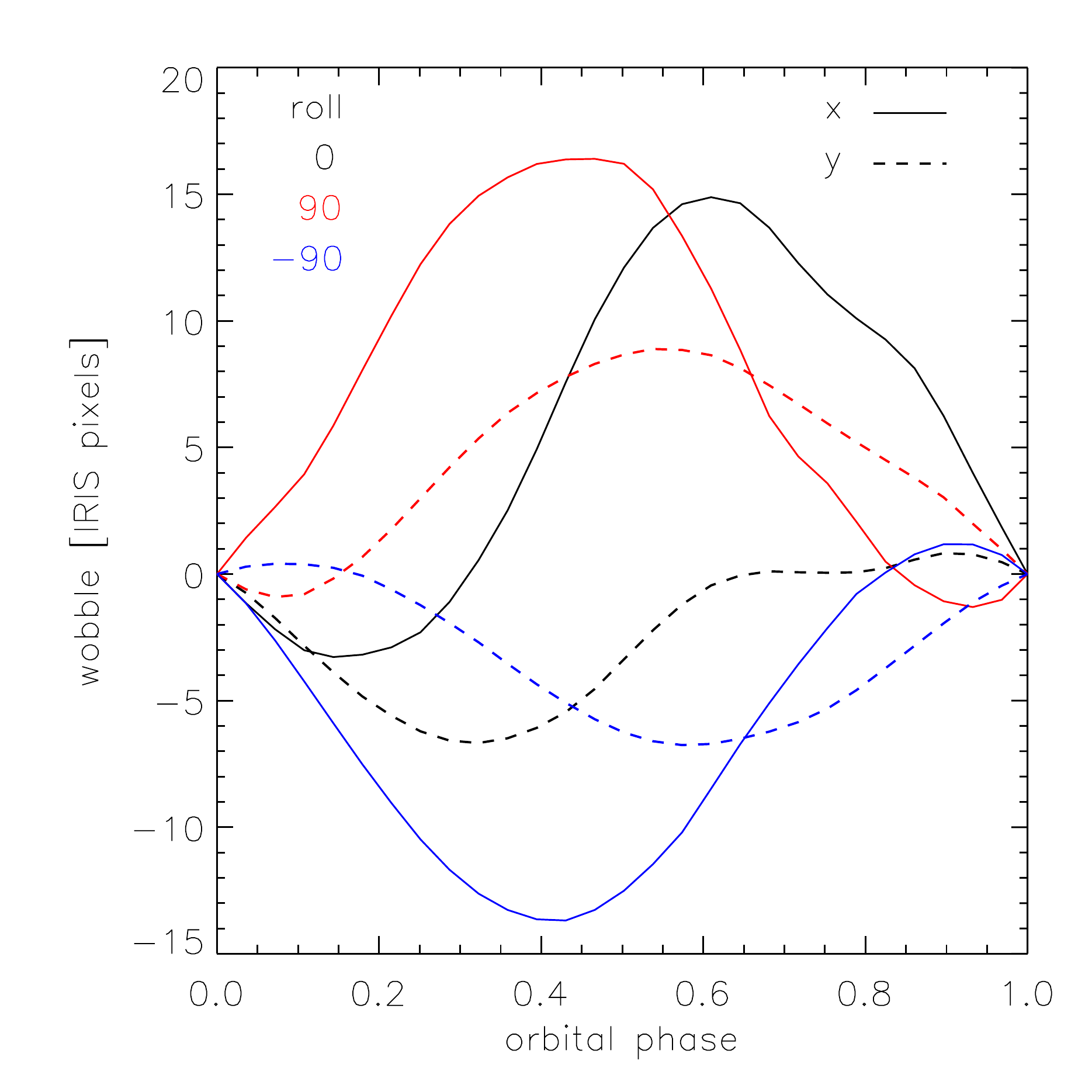}
              \includegraphics[width=0.5\textwidth,clip=]{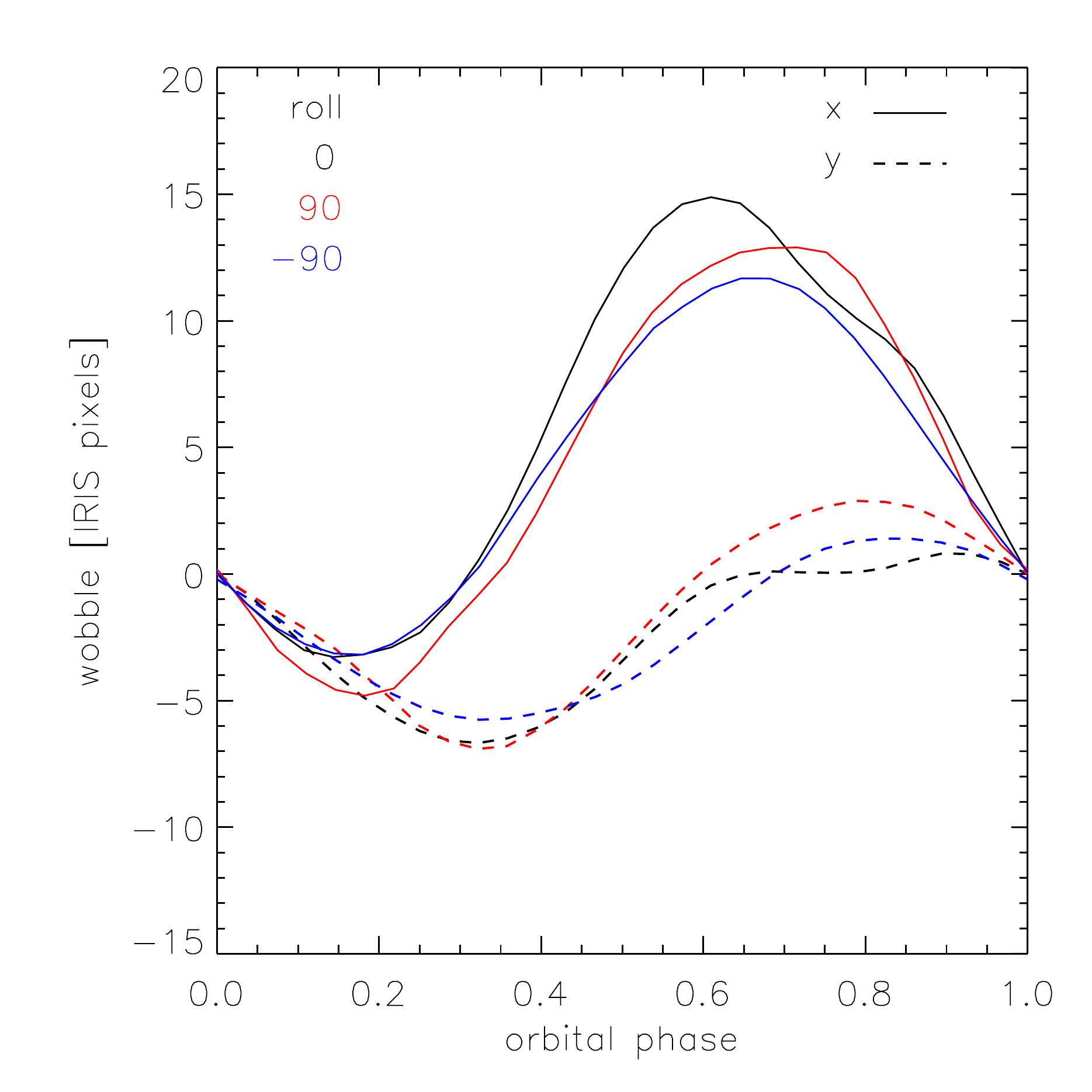}}
              \caption{Orbital wobble correction. Left panel:
IRIS orbital wobble in units of IRIS pixels, for the $x$- (solid lines)
and $y$ (dashed lines)-directions, and three roll angle values: 0
(black), +90 (red) and -90 (blue). Right panel:
Same curves as in the left panel, but with a phase shift of the wobble curves
for +90 and -90, of +0.25 and -0.25 respectively. The shifted curves
are very similar to the wobble curves for $0^\circ$ roll
angle. The orbital phase is zero at the time when IRIS passes through its
ascending node.}
   \label{fig_wobble}
   \end{figure}

Thermal flexing between the guide telescope and the main IRIS
telescope introduces a wobble in the observations on timescales of an
orbit (see Section~\ref{GTISS-instrument}).
We measured this wobble on-orbit by taking SJI 2832 channel data at the North Pole
and East Limb, at 20\,seconds cadence, covering more than an orbit, and then
running a cross-correlation algorithm on the data. To explore the
effect of roll angle on the wobble, we have carried out this analysis
for five roll angle values: 0$^\circ$, +90$^\circ$, -90$^\circ$, +45$^\circ$, -45$^\circ$.
The results of this analysis show that the wobble is, peak to peak
within an orbit, of the order of three arcsec in the $x$-direction, and one
arcsec in the $y$-direction (see Figure~\ref{fig_wobble}a, showing the wobble curves
for roll 0$^\circ$, +90$^\circ$, and -90$^\circ$).
For different roll angles we find that the wobble is of the same order
of magnitude as for 0$^\circ$ roll angle, although with a phase shift. In
particular we find that for roll angle $\alpha$, the wobble
can be approximated by the wobble for $0^\circ$ roll angle, shifted in phase
by  $\alpha/360^{\circ}$. This is illustrated in the right panel of Fig~\ref{fig_wobble}, where we
show the wobble curves of +90 and -90 shifted in phase by +0.25 and
-0.25 respectively, plotted over the wobble curves for 0$^\circ$ roll angle.

The effect of the wobble is corrected in regular IRIS operations with
orbital wobble tables (OWT), used to apply an adjustment to the
pointing which is compensating for the orbital wobble.
We estimate that the use of the OWTs corrects most of the wobble, with
a possible residual two IRIS pixel effect over the course of a full
orbit ($\approx\,97$ minutes). This residual is larger during
eclipse season.
For the angles used more frequently (0$^\circ$, $\pm 90^\circ$, $\pm 45^\circ$) we use
wobble tables derived by corresponding calibration data as described
above. However, deriving OWT for a larger set of roll angles is
unreasonably time-consuming (each angle requires calibration data for
two orbits, {\it i.e.}, 195 minutes, for each of the two pointings), especially
considering that the wobble calibrations will need to be run regularly,
several times a year. Therefore, for all other roll angles, we apply
the $0^\circ$ roll-wobble curves, shifted in phase by $\alpha/360^{\circ}$ as
described above. 

The availability of a continuous full-disk archive of UV and EUV
(SDO/AIA) and continuum (SDO/HMI) images at $\approx\,0.5$ arcsec spatial
resolution and $\approx24$\,seconds temporal resolution allows for the registration
of every IRIS slit-jaw image by means of cross-correlation of each
image with a field-of-view-matched, extracted sub-field from the nearest in
time SDO image of the appropriate waveband with comparable response. 
This technique will be used to cross-correlate
IRIS 1400 slit-jaw images with AIA 1700 image sub-fields, and IRIS
2832 slit-jaw images with HMI continuum image sub-fields, to extract
residual translation offsets for the IRIS slit-jaw images, which in
turn can be used to update the relevant World Coordinate System (WCS) keywords in the file
headers of both the slit jaw images and spectra, and thus improve the absolute accuracy of the
pointing information. 

\subsection{Wavelength Calibration}

The wavelength calibration of IRIS aims to remove instrumental
effects from the spectra so that these can be used to diagnose solar
conditions. The calibration involves two aspects, geometric correction
and relative calibration to photospheric spectral lines, which are
described in the following.

\subsubsection{Geometric Correction}
As with many spectrographs, IRIS suffers from slight spectral distortion
and misalignment, mainly rotation, spectral curvature, and the non-linearity of wavelength.  For analysis purposes
it is highly desirable to have the spectral and spatial coordinates
rectilinear with the grid of pixels and to have a constant dispersion
per pixel. 

 The geometry of IRIS is determined from heavily averaged, high
 signal-to-noise spectra, such as those created at the intermediate
 stage during the spectrograph flat fields.  From these average
 spectra the centers of spectral lines and fiducial marks are
 determined by Gaussian fitting.  The ``desired'' positions of the
 spectral lines and fiducials are calculated using the average line
 position at the center of the CCD.  The transformation to rectilinear
 coordinates is determined by fitting a two dimensional polynomial to
 the measured and ``desired'' line centers.

Once the geometric transformation is successfully determined, the
solution of the wavelength can be calculated.  The center positions of
neutral and low-ionization lines are used to calculate the
relationship between wavelength and pixels.  This solution is used to
modify the geometric result. An interactive tool
(\textsf{iris\_spec\_cal.pro}) has been implemented in SolarSoft to make this procedure relatively simple.

The geometry correction is implemented in \textsf{iris\_prep.pro}
(SolarSoft) using cubic interpolation, which causes some smoothing of
the resulting spectrum.  The spatial and spectral coordinate for every
pixel in the untransformed data is also available. 



\subsubsection{Relative Calibration}

In addition to the geometric corrections, there are also slight
thermal drifts and associated wavelength shifts. The low-Earth,
Sun-synchronous orbit of IRIS creates a thermal environment in which periodic temperature variations occur within the instrument on
timescales of one orbit (as well as over a year). To compensate
for some of these variations heaters are used to control some of the temperatures
within the instrument. The end result is a relatively stable thermal
environment that nevertheless shows some variations over an
orbit. 

Preflight thermal testing indicates that the high-order aspects of the
spectral geometry do not change, but the spectral lines shift with
thermal variation. These small drifts of the
spectral lines on the detector are of order two spectral pixels, {\it i.e.},
$\approx$\,6\,km~s$^{-1}$. During calibration these drifts were
determined by analyzing the wavelength shifts of neutral (and Fe {\sc ii}) lines, formed
in the photosphere or chromosphere, in the FUV 1, FUV 2, and NUV
detectors. These lines, when averaged in time and space, show very
little variation, compared to the large velocities in the upper
chromosphere and transition region, and can thus be used to determine
whether wavelength shifts for the differrent detectors are well
correlated with one another and/or with instrument temperatures.

\begin{figure}    
  \centerline{\includegraphics[width=1\textwidth,clip=]{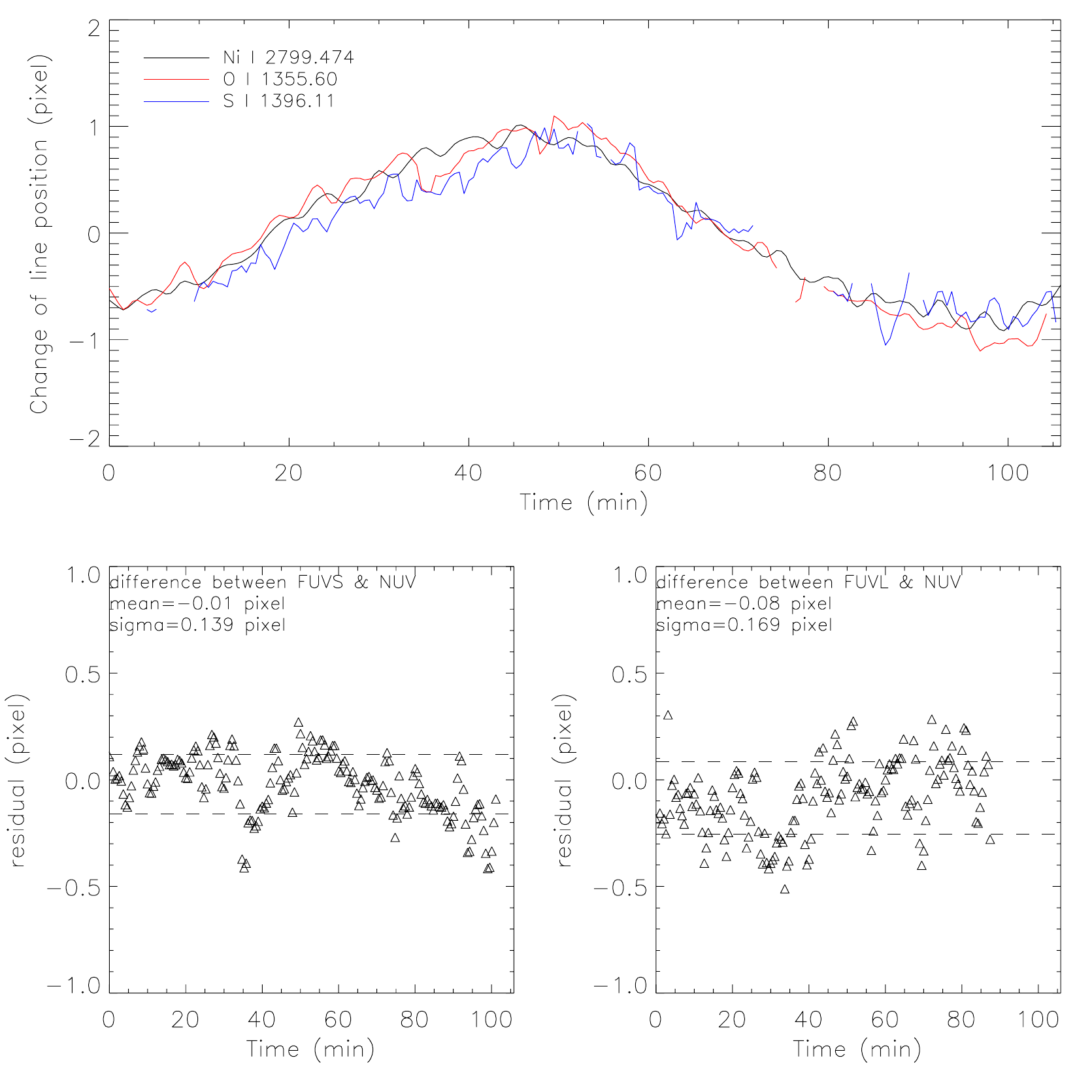}}
              \caption{Wavelength calibration approach. Top panel: correlation between wavelength shifts of
                neutral lines in FUV 1 (O {\sc i} 1355.6\,\AA), FUV 2 (S I
                1396.1\,\AA) and NUV (Ni {\sc i} 2799.5\,\AA) as a function of
                time during one orbit. Bottom panels:
                differences of wavelength shifts between FUV 1 and NUV
                (left panel) and between FUV
              2 and NUV (right panel).}
   \label{fig_wave_hui}
   \end{figure}

We did not find reliable correlations between the measured
temperatures and the observed wavelength
shifts. Therefore, it is not possible to derive a simple relationship between changes of
temperatures and line positions that can be applied to all
observations. However, we found that the wavelength shifts over the course
of an orbit are well correlated between FUV1, FUV2, and NUV after
removal of the orbital velocity of IRIS with respect to the Sun. The
wavelength shifts with time of the O {\sc i} 1355.6\,\AA\ line in FUV 1, S {\sc i} 1396.11\,\AA\ in FUV 2, and Ni {\sc i} 2799.47\,\AA\ line in NUV are well
correlated, as shown in Figure~\ref{fig_wave_hui}. The one-$\sigma$ residual of the
difference between the temporal variations of NUV with those in FUV is about 0.5 km~s$^{-1}$. This is much smaller
than the typical TR velocity [$\approx\,5$\,km~s$^{-1}$]. These correlations are
found to hold for different roll angles of the spacecraft [$\pm$ 90 degrees].

Using these results, the IRIS wavelength-calibration approach is to
determine, for every NUV spectrum taken, the wavelength shifts of the Ni I
2799.474\,\AA\ line. This line has significantly higher signal-to-noise (S/N) than the
FUV neutral lines. We average these shifts along the slit to determine the
average thermal drift of the line and then use this to establish a wavelength
calibration of not only the NUV passband, but also the FUV 1 and FUV 2
(using the correlations discussed above). 

Because the intrinsic average velocity exhibited by these photospheric
neutral lines is less than 0.5 km~s$^{-1}$ (based on MHD simulations), our
calibration leads to an absolute calibration of the NUV passband with a precision of $\approx1$
km~s$^{-1}$. 

\subsection{Throughput}


\subsubsection{FUV Passbands}

IRIS's off-limb capability and high sensitivity allow us to use
UV-bright stars as calibration sources. We have identified about a
dozen stars that should be detectable by IRIS. 
In August 2013, we demonstrated our stellar calibration capability with early
observations of HD\,86360 and HD\,91316. Stellar calibration allows us to
 convert IRIS fluxes into physical units and to measure, on-orbit,
 the effective area of the SJIs and spectrograph. By observing the
 same stars throughout the mission we will have an excellent measure
 of the change in throughput due to contamination and degradation.

\begin{table}[h]
  \caption{Effective area for each IRIS science channel}
   \begin{tabular}{llll}
      Channel & Effective Area [cm$^2$] & Effective Area [cm$^2$] &Radiometric conversion\\ 
 & Prelaunch & Flight &
 [(erg\,s$^{-1}$\,sr$^{-1}$cm$^{-2}$\AA$^{-1}$)\\
 & & & /(DN\,s$^{-1}$)]\\
\hline
      SJI 1330\,\AA &  0.5  & 0.41&N/A\\
      SJI 1400\,\AA &  0.6  & 0.81&N/A\\
      SJI 2796\,\AA &  0.005 & N/A&N/A\\
      SJI 2832\,\AA &  0.004 & 0.0043&N/A\\
      NUV SG        &  0.2 & N/A& N/A\\
      FUV 1 SG        & 1.6 &1.2 &2960 \\
      FUV 2 SG        & 2.2 &2.2 &1600\\
  \end{tabular}
\label{table_eff2}
\end{table}

The August 2013 campaign included specifically designed observing
programs for two UV-bright stars: HD\,86360 and HD\,91316. We have
successfully observed both stars in the SJI 1400 and 1330 filters. 
By changing pointings and slit locations with time, the FUV spectrum of
HD\,91316 was successfully recorded at about 20 locations of the slit.

We have evaluated the effective areas for both SJI and SG, and
performed radiometric calibration for the FUV spectra of IRIS.
Spectra of both stars were observed by the {\it International Ultraviolet Explorer} (IUE) about 30 years ago. The IUE spectral radiances are given with uncertainties of 10\,--\,15\% (one $\sigma$). Implicit in cross calibration with stars is the assumption that the irradiances of the stars are nearly constant in time, which is the case for both of these stars. 

We have evaluated the effective areas of the SJI 1400 and 1330 band passes
using data for both stars. The average effective area is about 35\,\%
larger than the pre-flight value for the SJI 1400 filter 
and 12\,\% smaller than the pre-flight value for the SJI 1330 filter. 

Using data from the star HD\,91316, we have evaluated the effective areas
of the two FUV wavelength bands. The average effective area is about
1.2 cm$^2$ for FUV 1 and 2.2 cm$^2$ for FUV 2. Both are within the
error margin of the pre-flight estimates.

We have performed radiometric calibration of the FUV spectra of IRIS
based on the HD\,91316 observation.  By comparing the calibrated
IUE flux with the uncalibrated IRIS spectrum of the star, we have
derived the radiometric calibration factor for IRIS. This factor is
about 2960 in the FUV 1 band and 1600 in the FUV 2 band. Using this
factor, we can convert the observed countrate (DN\,s$^{-1}$) into radiance of
the spectrum (erg\,s$^{-1}$\,sr$^{-1}$\,cm$^{-2}$\,\AA$^{-1}$). These values are summarized in Table~\ref{table_eff2}.

\subsubsection{NUV Passbands}

The stars we observed are not bright enough in the NUV. For the
calibration of the NUV channels, we compared IRIS quiet-Sun
disk-center observations on 19\,Jul\,2013 with full-Sun
spectra of 15\,Jul\,2013 taken with the {\it Solar Stellar
  Irradiance Comparison Experiment} (SOLSTICE) onboard the {\it Solar Stellar
  Irradiance Comparison Experiment} (SORC). We scaled the IRIS data to full disk, and
applied a limb darkening correction
$F_{\rm tot}/I_{\rm ctr}=0.588$ at 2800\,\AA \cite{Allen64}.
The SOLSTICE spectrum was then folded with the IRIS pre-launch effective
area. A comparison of the spectra shows a good match in the wings of
the Mg {\sc ii} h and k lines, with discrepancies near the line core. The latter is to be
expected since the line cores are affected by solar activity: the SOLSTICE
spectrum includes all active regions on the disk, while IRIS only observed
at the quiet Sun near disk center.
For the wing of the lines, we can nevertheless determine how well our pre-flight effective area values
for the NUV 2832 SJI channel (in the wings of Mg {\sc ii} h/k) compare with the
flight values. We found that the
flight values are 8\% higher than the pre-flight values, {\it i.e.}, within
the uncertainty of the calibration method.

The NUV 2796 SJI channel and the NUV spectrograph
pre-flight values are given in Table~\ref{table_eff2}.

\subsubsection{Temporal Evolution}

To determine the long-term evolution of the sensitivity of the various
IRIS passbands, we will continue to perform stellar calibrations as
the targets appear in the IRIS field of view. This will help calibrate
any long-term trends of the sensitivity of the FUV channels. In addition, IRIS obtains observations of a quiet region near disk center on a
daily basis in order to monitor throughput of both FUV and NUV
channels.

\section{Data Processing} 
      \label{S-data} 

\subsection{Level 0 to 1.5}

IRIS data processing leverages the heritage infrastructure developed
for SDO and {\it Hinode}. Raw science data flows from the Mission Operation
Center (MOC) at NASA Ames to the Joint Science Operations Center
(JSOC) at Stanford University and LMSAL. The IRIS data system uses the
same infrastructure as that for SDO/AIA and HMI. At the JSOC, 
the science telemetry is archived, with a copy exported to the backup
site at LMSAL. After receipt from the MOC, the telemetry is converted
into Level 0 image files (see Table~\ref{table_data}). To generate
Level 1 data, mirroring of various axes is removed and various headers
are added containing information about temperatures, roll, and
pointing. It is automatically produced in a pipeline at the JSOC. 
All these data levels are stored within the JSOC data system.

Level 1.5 is produced at LMSAL and is the lowest level of
scientifically-useful data. Dark current and systematic (pedestal) offsets are
removed, and flat-fielding corrections for
telescope and CCD properties, as well as geometric and wavelength
corrections, are applied. All images are
mapped to a common spatial plate scale. Spectral images are also
remapped to align with an equal-sized array where wavelength and
spatial coordinates align with the grid. An array mapping the
wavelength axis to physical wavelength is created in this process. As
with AIA, equivalent procedures to those used internally to transform
Level 1 to Level 1.5 are distributed via SolarSoft as \textsf{iris\_prep.pro}. 

Level 1 and 1.5 data products are still organized per “CCD frame”,
{\it i.e.}, every exposure taken by IRIS is stored as a FITS file.
The format of these files is compressed FITS files, similar to AIA and
HMI: full frames are stored even when only some regions of interest (ROI)
were obtained during a particular exposure. In this case, most of the
FITS array is set to a ``Not a Number'' (NaN), with only the ROI
displaying observed data numbers.

  \begin{table}
   \caption{IRIS Data Levels}
   \label{table_data}
    \begin{tabular}{ll}     
      \hline                   
      {\bf Level} & {\bf Description}\\
\hline
0 &depacketized,  raw images with housekeeping and overscans\\
1 & reorient images to common axes: north ``up'' (0$^\circ$ roll), increasing $\lambda$/x to right\\
1.5 & dark current and offsets removed \\
& flag bad pixels and spikes pixels \\
&flat field correction\\
& geometric and wavelength calibration\\
1.6 & physical units (exposure and photon conversion)\\
{\bf 2} &{\bf recast as rasters and SJI timeseries, standard science product}\\
3 &recast as 4D cubes for NUV/FUV and 3D cubes for SJI for CRISPEX\\
HCR & Description of observing sequences, ingested by HCR at LMSAL\\ 

   \end{tabular}
 \end{table}

\subsection{Level 2 and 3}

Level 2 is generated at LMSAL using Level 1.5 data. Level 3 data can be
produced by the user using \textsf{iris\_make\_fits\_level3.pro} (SolarSoft).
Level 2/3 data products group individual ``CCD frames'' (Level 1.5) into
logical units. There are two different types of Level 2 data products: spectral
rasters and slit-jaw image timeseries. These are the data products that will be most widely used. 
As calibration procedures improve, these may be regenerated as appropriate. 

\subsubsection{Level 2 Spectral Rasters}

Spectral rasters are defined as a series of spectral frames that are:  
\begin{itemize}
\item within one execution of an OBS list
\item a contiguous set with identical sgn($\delta$H), in which $\delta$H is the value of the perpendicular displacement of the active secondary (controlled by PZT ABC at the FRM and OBS list level)
\item if $\delta$H = 0 ({\it i.e.}, sit-and-stare), then all spectra go into one (timeseries) file
\end{itemize}

All spectral regions of interest of one raster step are stacked into
one ``raster'' file, including
NUV and FUV ({\it i.e.}, unlike Level 1 these different spectra are now
stored into one file). The spectral windows are stacked as extensions (as part of the new FITS 3 standard).
Each spectral region within a CRS is stored into a separate
extension. 
Each extension has auxiliary header info that describes time,
position, etc.

\subsubsection{Level 2 Slitjaw Images}

All slitjaw (SJI) images of one execution of an OBS list and for one
filterwheel (FW) position and readout region (CRS) are stored as one
separate file. These SJI files contain a time series. There are separate SJI files for each filterwheel choice and for each different CRS region.

\subsubsection{Level 3}

Level 3 data groups the Level 2 data into products that allow easy
quicklook and analysis using the IRIS-specific \textsf{crispex.pro} quicklook
tool \cite{crispex}. In practice this means that timeseries of spectral rasters are re-ordered into 4D data cubes (FITS-files).  

\textsf{CRISPEX} was developed to analyze data taken with the CRISP Fabry-Perot
interferometer \cite{crisp}. This instrument takes two-dimensional images of the Sun within a narrow bandpass centered at a specific wavelength. CRISP is usually used to perform rapid wavelength scans through a chromospheric line. To facilitate speedy access to the large 4D (x, y, $\lambda$, t) datasets \textsf{CRISPEX} does not read the data into memory, but reads it from disk as it is accessed. To optimize the speed at which the data is displayed, it is stored in two separate formats: x, y, $\lambda$, t and $\lambda$, t, x, y. This allows for rapid display of both images, $\lambda$-intensity, $\lambda$-t and $\lambda$-x plots, allowing easy quicklook access.

To fully exploit \textsf{CRISPEX}, IRIS Level 2 data is rendered into CRISPEX-ready files. The latter are the IRIS Level 3 data products.

In particular, a Level 3 IRIS \textsf{CRISPEX} file will be based on a specific
execution of an OBS list and contain either of these:
\begin{itemize}
\item one IRIS (synthetic) spectral window stored in {\it
    xy$\lambda$t} or {\it$\lambda$txy} format
\item all spectral windows stored in {\it xy$\lambda$t} or {\it $\lambda$txy} format
\end{itemize}

Level 3 data is not distributed publicly because the data volume is
very large, but can be produced from Level 2 data
using SolarSoft (\textsf{iris\_make\_fits\_level3.pro}).

\subsection{Other Products}

Browse products are generated from Level 2 and 3 images. Heliophysics
Coverage Registry (HCR) products are
derived from the planned operations and the resulting data, and are
recorded in the Heliophysics Coverage Registry, which is part of the
Heliophysics Event Knowledgebase \cite{Hurlburt12}. 

\subsection{Data Availability}

Data from IRIS are made available to the community through several
paths. Level 2 data products are the standard science products. These
are available through the IRIS website
(\href{http://iris.lmsal.com}{iris.lmsal.com}) in a manner similar to
{\it Hinode}, TRACE and the AIA
cutout web services.  Level 2 data are also available at the
{\it Hinode} archive at the University of Oslo. Higher-level descriptions of the data products
are maintained within the Heliophysics Coverage Registry and solar
events associated with these data products are captured within the
Heliophysics Event Registry. IRIS data are available via the Virtual
Solar Observatory \cite{vso} through a variety of avenues.

Level 1 data are available directly from the JSOC using the same
procedures as developed for SDO. These Level 1 data are a published series that is either available through the JSOC subscription mechanisms or through the lookdata web interface. 

\section{Numerical Modeling} 
      \label{S-sims} 

\subsection{Overview}
The IRIS investigation includes an extensive radiative-MHD
numerical simulation component to enable full forward-modeling of the domain from the top
of the convective envelope to the low corona. Such a comprehensive
approach to numerical modeling is critical for a full understanding of
the interface region and the diagnostics that cover this
region. Advances in algorithms, supercomputing capability, and
parallelization techniques now allow such computations at a spatial
and temporal resolution that is adequate for comparisons with IRIS
observables, including various physical mechanisms that have a
critical role in the thermodynamic evolution of the low solar
atmosphere.

The rapid advances in computational infrastructure and algorithms
have revolutionized the state-of-the-art of numerical simulations and
the role these play in interpreting observations. 
These advances are dramatically reducing the
idealizations and simplifications that are made in order to enable the
computations. As part of the pre-launch IRIS science investigation,
the co-I team has worked towards improving the realism of various
numerical models through intensive collaboration and annual
workshops. 

These efforts have been focused on:
\begin{itemize}
\item increased spatial resolution,
\item wide range of magnetic-field configuration, including
  small-scale flux emergence,
\item increased volume of numerical domain,
\item inclusion of non-equilibrium ionization of hydrogen and helium,
\item inclusion of effects of ion--neutral interactions in the partially
  ionized chromosphere,
\item inclusion of kinetic effects through a hybrid PIC/MHD approach,
\item line formation and interpretation of C {\sc ii} 1335\,\AA\ and Mg {\sc ii} h/k 2796/2803\,\AA,
\item comparison of physical and diagnostic results from various
  numerical models.
\end{itemize}

Despite the advances in numerical modeling, it is clear that
simplifying asumptions are still required to enable the
calculations. The IRIS modeling philosophy is that comparisons between
IRIS (and other) observations and synthetic diagnostics from numerical
models are used to determine which physical mechanisms reduce any
discrepancies between observations and models, and which assumptions
are too idealized and should be relaxed. For example, preliminary
studies of the formation of the Mg {\sc ii} h and
k lines indicate that they provide a sensitive diagnostic for upper
chromospheric heating and turbulence, and that current models lack the
right amount of heating and turbulence just below the transition
region, likely because time-dependent hydrogen ionization and
ambipolar diffusion have not yet been fully taken into account, and/or
because the simulations lack spatial resolution. This combined
approach, in which observations and models are continuously compared,
guiding the next development steps for the models,
is a major part of the IRIS science investigation.

The numerical modeling team of the IRIS science investigation has
played a crucial role throughout the development of the IRIS
instrument requirements, for example by assisting with assessing the impact of
telescope and instrument design choices in terms of full width half
maximum of the Solc filter, or studying the requirements for spectral
resolution in the NUV SG passband. IRIS modeling has helped not only
with determining requirements but also assisted in design choices later
in the development program. The IRIS models have also been highly
valuable in assisting with prelaunch optimization of observing
sequences, through the use of the so-called IRIS simulator which
allows a user to observe the models with the same observing tables
that IRIS uses. This code is available in the IRIS SolarSoft tree.￼

\subsection{Publicly Available Models and Papers}

To assist the community with the interpretation of IRIS data, several
numerical models computed with the \textsf{Bifrost} code from the University of
Oslo are for the first time being publicly released (see \href{http://sdc.uio.no/search/simulations}{sdc.uio.no/search/simulations}). The numerical models in
question have also been used to investigate the line formation of Mg
{\sc ii} h and k \cite{Leenaarts13a}, the relationship between spectral
features in the Mg {\sc ii} h/k lines and physical variables in the
chromosphere \cite{Leenaarts13b}, how IRIS observations of Mg {\sc ii}
h/k can be used to diagnose chromospheric conditions
\cite{Pereira13}, the line formation of the C {\sc ii} 1335\,\AA\ lines, 
and the effects of non-equilibrium ionization on the
IRIS FUV diagnostics \cite{Olluri13}.

  \begin{table}
   \caption{IRIS Documentation and URLs}
   \label{table_www}
    \begin{tabular}{ll}     
      \hline                   
      {\bf Description} & {\bf URL}\\
\hline
Main website & \href{http://iris.lmsal.com}{iris.lmsal.com}\\
\hline
Operations & \href{http://iris.lmsal.com/operations.html}{iris.lmsal.com/operations.html}\\
\hline
Data search & \href{http://iris.lmsal.com/search/}{iris.lmsal.com/search/} \\
&\href{http://sdc.uio.no/search/form}{sdc.uio.no/search/form}\\
\hline
Recent observations &{\tiny
\href{http://www.lmsal.com/hek/hcr?cmd=view-recent-events\&instrument=iris}{www.lmsal.com/hek/hcr?cmd=view-recent-events\&instrument=iris}} \\
\hline
IRIS Today & \href{http://iris.lmsal.com/iristoday/}{iris.lmsal.com/iristoday}\\
\hline
Documentation & \href{http://iris.lmsal.com/documents.html}{iris.lmsal.com/documents.html}\\
\hline
  \end{tabular}
 \end{table}

\section{Conclusion} 
      \label{S-conclusions} 

The IRIS door was opened and the first images and spectra were acquired on
17 July 2013. Since then, IRIS has successfully finished its initial 60 day
observing plan, which was focused on obtaining a variety of datasets
(sit-and-stare, dense rasters, coarse rasters) for a range of different
targets (quiet Sun, active regions, coronal holes, prominences). 
The IRIS slit-jaw images and spectra reveal a variety of highly
dynamic and finely scaled structures, often involving high
velocities. Both the FUV and NUV channels of IRIS perform very well,
with C {\sc ii}, Si {\sc iv}, Mg {\sc ii} k, and Mg {\sc ii} h spectra providing a novel window
on the intimate coupling between the photosphere, chromosphere, and transition region.
Hotter active regions and/or flares also show Fe {\sc xii} and Fe {\sc xxi}
spectra, extending IRIS's temperature coverage well into the hot
corona. The spatial, spectral and temporal resolution are excellent,
reaching 0.33\,--\,0.4 arcsec, $<1$ km~s$^{-1}$ and down to 1.5\,seconds,
respectively. The various channels are well co-aligned and the image
stabilization system is excellent. Typical observing sequences take
spectra and images at cadences between 3 and 30s. Calibration work
to monitor long-term behavior continues, but calibrated datasets are
now available to the public without restriction within a few days of
observations. Extensive documentation on operations, calibration, data
analysis, and numerical simulations is available in the form of
so-called IRIS Technical Notes (ITNs) on the \href{http://iris.lmsal.com}{iris.lmsal.com}
website, as outlined in Table~\ref{table_www}.

\appendix

\section{Observing Tables}

The default observing tables are given in Tables~\ref{table_default1},
\ref{table_default2}, \ref{table_default3}.

  \begin{table}[h]
   \caption{IRIS default tables: basic raster modes}
   \label{table_default1}
    \begin{tabular}{llll}     
      \hline                   
      {\bf OBS number} & {\bf Raster step} & {\bf Raster size} & {\bf Description}\\
      & [arcsec] & [arcsec$^2$] & \\
\hline
1&0.33&0.3$\times$30&Small sit-and-stare\\
2&0.33&0.3$\times$60&Medium sit-and-stare\\
3&0.33&0.3$\times$120&Large sit-and-stare\\
4&0.33&0.3$\times$175&Very large sit-and-stare\\
5&0.33&0.33$\times$30&Small dense 2-step raster\\
6&0.33&0.33$\times$60&Medium dense 2-step raster\\
7&0.33&0.33$\times$175&Very large dense 2-step raster\\
8&1&1$\times$60&Medium sparse 2-step raster\\
9&1&1$\times$120&Large sparse 2-step raster\\
10&1&1$\times$175&Very large sparse 2-step raster\\
11&2&2$\times$60&Medium coarse 2-step raster\\
12&2&2$\times$120&Large coarse 2-step raster\\
13&2&2$\times$175&Very large coarse 2-step raster\\
14&0.33&1$\times$30&Small dense 4-step raster\\
15&0.33&1$\times$60&Medium dense 4-step raster\\
16&0.33&1$\times$175&Very large dense 4-step raster\\
17&1&3$\times$60&Medium sparse 4-step raster\\
18&1&3$\times$120&Large sparse 4-step raster\\
19&1&3$\times$175&Very large sparse 4-step raster\\
20&2&6$\times$60&Medium coarse 4-step raster\\
21&2&6$\times$120&Large coarse 4-step raster\\
22&2&6$\times$175&Very large coarse 4-step raster\\
23&0.33&2.32$\times$30&Small dense 8-step raster\\
24&0.33&2.32$\times$60&Medium dense 8-step raster\\
25&0.33&2.32$\times$175&Very large dense 8-step raster\\
26&1&7$\times$60&Medium sparse 8-step raster\\
27&1&7$\times$120&Large sparse 8-step raster\\
28&1&7$\times$175&Very large sparse 8-step raster\\
29&2&14$\times$60&Medium coarse 8-step raster\\
30&2&14$\times$120&Large coarse 8-step raster\\
31&2&14$\times$175&Very large coarse 8-step raster\\
32&0.33&5$\times$60&Medium dense 16-step raster\\
33&0.33&5$\times$120&Large dense 16-step raster\\
34&0.33&5$\times$175&Very large dense 16-step raster\\
35&1&15$\times$60&Medium sparse 16-step raster\\
36&1&15$\times$120&Large sparse 16-step raster\\
37&1&15$\times$175&Very large sparse 16-step raster\\
38&2&30$\times$120&Large coarse 16-step raster\\
39&2&30$\times$175&Very large coarse 16-step raster\\
40&0.33&20.8$\times$120&Large dense 64-step raster\\
41&0.33&20.8$\times$175&Very large dense 64-step raster\\
42&1&63$\times$120&Large sparse 64-step raster\\
43&1&63$\times$175&Very large sparse 64-step raster\\
44&2&126$\times$120&Large coarse 64-step raster\\
45&2&126$\times$175&Very large coarse 64-step raster\\
46&0.33&131.7$\times$175&Very large dense raster\\
47&0.33&31.35$\times$175&Dense synoptic raster\\
48&1&35$\times$175&Sparse synoptic raster\\
49&2&34$\times$175&Coarse synoptic raster\\
\hline
    \end{tabular}
 \end{table}

  \begin{table}[h]
   \caption{IRIS default tables: SJI, exposure times, rebinning}
   \label{table_default2}
    \begin{tabular}{rl}     
      \hline                   
      {\bf OBS number} & {\bf Description}\\
      
\hline
0 & C {\sc ii} Si {\sc iv} Mg {\sc ii} h/k Mg {\sc ii} w\\
100 & C {\sc ii} Si {\sc iv} Mg {\sc ii} h/k Mg {\sc ii} w s\\
200 & C {\sc ii} Si {\sc iv} Mg {\sc ii} w s\\
300 & C {\sc ii} Mg {\sc ii} h/k Mg {\sc ii} w s\\
400 & Si {\sc iv} Mg {\sc ii} h/k Mg {\sc ii} w s\\
500 & C {\sc ii} Mg {\sc ii} w s\\
600 & Si {\sc iv} Mg {\sc ii} w s\\
700 & Mg {\sc ii} h/k Mg {\sc ii} w s\\
800 & Si {\sc iv} Mg {\sc ii} h/k Mg {\sc ii} w\\
900 & C {\sc ii} Mg {\sc ii} h/k Mg {\sc ii} w\\
1000 & C {\sc ii} Si {\sc iv} Mg {\sc ii} w\\
1100 & C {\sc ii} Si {\sc iv} Mg {\sc ii} h/k\\
1200 & C {\sc ii} Si {\sc iv}\\
1300 & C {\sc ii} Mg {\sc ii} h/k\\
1400 & Si {\sc iv} Mg {\sc ii} h/k\\
1500 & C {\sc ii}\\
1600 & Si {\sc iv}\\
1700 & Mg {\sc ii} h/k\\
1800 & Mg {\sc ii} w\\
1900 & Mg {\sc ii} h/k Mg {\sc ii} w\\
\hline
 0& Exp time 1s\\
2000& Exp time 0.5s\\
4000& Exp time 2s\\
6000& Exp time 4s\\
8000& Exp time 8s\\
10\,000& Exp time 15s\\
12\,000& Exp time 30s\\
\hline
 0& Spatial$\times$1, Spectral$\times$1\\
20\,000& Spatial$\times$1, Spectral$\times$2\\
40\,000& Spatial$\times$1, Spectral$\times$4\\
60\,000& Spatial$\times$1, Spectral$\times$8\\
80\,000& Spatial$\times$2, Spectral$\times$1\\
100\,000& Spatial$\times$2, Spectral$\times$2\\
120\,000& Spatial$\times$2, Spectral$\times$4\\
140\,000& Spatial$\times$2, Spectral$\times$8\\
160\,000& Spatial$\times$4, Spectral$\times$1\\
180\,000& Spatial$\times$4, Spectral$\times$2\\
200\,000& Spatial$\times$4, Spectral$\times$4\\
220\,000& Spatial$\times$4, Spectral$\times$8\\
\hline
 0& FUV spectrally rebinned$\times$2\\
250\,000& FUV spectrally rebinned$\times$2\\
500\,000& FUV spectrally rebinned$\times$4\\
750\,000& FUV spectrally rebinned$\times$8\\
\hline
\end{tabular}
\end{table}

 \begin{table}[h]
   \caption{IRIS default tables: SJI cadence, compression, linelists}
   \label{table_default3}
    \begin{tabular}{rl}     
      \hline                   
      {\bf OBS number} & {\bf Description}\\
      
\hline
 0 & SJI cadence 10s (60s for slow) \\
1\,000\,000 & SJI cadence 0.25$\times$ faster\\
2\,000\,000 & SJI cadence 0.5$\times$ faster\\
3\,000\,000 & SJI cadence 3$\times$ faster\\
4\,000\,000 & SJI cadence 10$\times$ faster\\
\hline
 0 & Default lossy compression \\
10\,000\,000 & Lossless compression \\
\hline
 0 & Large Linelist \\
20\,000\,000 & Medium Linelist\\
40\,000\,000 & Small Linelist\\
60\,000\,000 & Flare Linelist\\
80\,000\,000 & Full readout\\
\hline
   \end{tabular}
 \end{table}

\begin{acks}
The effort required to build a mission such as IRIS requires a large,
skillful, and dedicated team. We wish to acknowledge many individuals
who contributed to the success of IRIS: Geoff Andrews, Nate Caditan, Brock
Carpenter, Jay Dusenbury, Cliff Evans, Chuck Fischer, Scott Green, George
Dankiewicz, Robert Honeycutt,
James Irwin, Harjeet Janda, Dwana Kacensky, Pete Kacensky, Mike
Marticorena, Mark Ridley, John Serafin, David Schiff, Richard Shine, Araya
Silpikul, Greg Slater, Shanti Varaitch, Leah Wang, Ross Yamamoto and Kent Zikuhr (Lockheed Martin). We are grateful to
Mats L\"ofdahl for providing his IDL phase-diversity code and teaching
us how to use it. We would like to thank Jeff Newmark, Joseph Davila,
and Adrian Daw at NASA for their support and encouragement. 
This work is supported by NASA under contract NNG09FA40C and the Lockheed
Martin Independent Research Program. The data downlink to Svalbard is
funded by the Norwegian Space Centre (NSC) through an ESA PRODEX
contract. We would like to thank Bo Andersen and Paal Brekke for their
efforts in making the NSC downlink support possible.
\end{acks}

  
\bibliographystyle{spr-mp-sola}

\bibliography{iris_mission}

\end{article} 

\end{document}